\documentclass[11pt,twocolumn]{aastex63}
\usepackage{url}
\usepackage{gensymb}
\usepackage{textcomp}
\usepackage{xspace}
\usepackage{natbib,amsmath} \usepackage{graphicx}
\usepackage{hyperref} \usepackage{float} \usepackage{subfigure}
\usepackage{tabularx}
\providecommand{\e}[1]{\ensuremath{\times 10^{#1}}}
\usepackage{bm}

\newcommand{\ang}{\AA\xspace}

\newcommand{\g}{\mathrm{g}}
\newcommand{\K}{\mathrm{K}}

\newcommand{\s}{\mathrm{s}}

\newcommand{\p}{\mathrm{p}}

\usepackage{array}

\begin{document}
\title{Escaping Helium from TOI 560.01, a Young Mini Neptune}

\correspondingauthor{Michael Zhang}
\email{mzzhang2014@gmail.com}

\author[0000-0002-0659-1783]{Michael Zhang}
\affiliation{Department of Astronomy, California Institute of Technology, Pasadena, CA 91125, USA}

\author[0000-0002-5375-4725]{Heather A. Knutson}
\affiliation{Division of Geological and Planetary Sciences, California Institute of Technology}

\author[0000-0002-6540-7042]{Lile Wang}
\affiliation{Center for Computational Astrophysics, Flatiron Institute, New York, NY 10010, USA}

\author[0000-0002-8958-0683]{Fei Dai}
\affiliation{Division of Geological and Planetary Sciences, California Institute of Technology}

\author[0000-0003-0563-0493]{Oscar Barrag\'an}
\affiliation{Sub-department of Astrophysics, Department of Physics, University of Oxford, Oxford, OX1 3RH, UK}

\begin{abstract}
We report helium absorption from the escaping atmosphere of TOI 560.01 (HD 73583b), a $R=2.8 R_\Earth$, $P=6.4$ d mini Neptune orbiting a young ($\sim$600 Myr) K dwarf.  Using Keck/NIRSPEC, we detect a signal with an average depth of $0.68 \pm 0.08$\% in the line core.  The absorption signal repeats during a partial transit obtained a month later, but is marginally stronger and bluer, perhaps reflecting changes in the stellar wind environment.  Ingress occurs on time, and egress occurs within 12 minutes of the white light egress, although absorption rises more gradually than it declines.  This suggests that the outflow is slightly asymmetric and confined to regions close to the planet.  The absorption signal also exhibits a slight 4 km/s redshift rather than the expected blueshift; this might be explained if the planet has a modest orbital eccentricity, although the radial velocity data disfavors such an explanation.  We use XMM-Newton observations to reconstruct the high energy stellar spectrum and model the planet's outflow with 1D and 3D hydrodynamic simulations.  We find that our models generally overpredict the measured magnitude of the absorption during transit, the size of the blueshift, or both.  Increasing the metallicity to 100$\times$ solar suppresses the signal, but the dependence of the predicted signal strength on metallicity is non-monotonic.  Decreasing the assumed stellar EUV flux by a factor of 3 likewise suppresses the signal substantially.
\end{abstract}


\section{Introduction}
\label{sec:introduction}
There is a growing body of evidence suggesting that atmospheric mass loss dramatically shapes the population of close-in exoplanets detected by transit surveys.  This population is dominated by planets with radii between $1-4$ R$_\Earth$ (`sub-Neptunes'), which have no solar system analogue.  Smaller sub-Neptunes ($1-1.7$ Earth radii) appear to have Earth-like bulk compositions and are commonly referred to as `super Earths', while larger sub-Neptunes ($2-3$ Earth radii) are called `mini Neptunes' and have low bulk densities indicating the presence of volatile-rich envelopes that typically constitute a few percent of their total mass.  The two populations are separated by a gap in the radius distribution where few planets reside \citep{fulton_2017,fulton_2018}.  Sub-Neptunes are challenging to characterize, but new telescopes, observing techniques, and numerical models are opening up this frontier to exploration.

It has been suggested that all sub-Neptunes were formed with hydrogen-rich envelopes, which were then stripped away from the most highly irradiated and least massive planets.  For young, low-density planets on close-in orbits, photoevaporation can drive strong hydrodynamic outflows that rapidly remove hydrogen-rich gas (e.g., \citealt{owen_2017, mills_2017}).  However, this mass loss may also be driven by the newly formed planet's own internal luminosity \citep{ginzburg_2018,gupta_2019}.  An alternate explanation for the radius gap is that it has nothing to do with mass loss, but is instead because cores have a broad mass distribution, with the smaller cores having never accreted gas in the first place \citep{lee_2021}.  It is also possible that some mini Neptunes have no hydrogen-rich envelopes at all, but instead formed with substantial water-rich envelopes (e.g. \citealt{mousis_2020}).  This could dramatically change the mass loss rate, especially that of helium, which would have been already lost to space alongside the primordial hydrogen.

Empirical measurements of mass loss from young sub-Neptunes provide a critical test of these competing hypotheses.  We expect that mass loss rates will be highest at relatively early times, when the star's high-energy flux is enhanced and the planet is still inflated \citep{owen_2018}.  By searching for evidence of outflows during this critical early period, we can determine whether or not a young sub-Neptune planet possesses a hydrogen and helium-rich envelope.  When an outflow is detected, it can be used to test and refine mass loss models, with the eventual goal of developing an accurate understanding of atmospheric evolution through the planet's life.

We can search for outflows by measuring the amount of absorption from hydrogen and/or helium during the transit, but the small size of mini Neptunes makes them challenging targets for this technique.  In a recent study, we obtained the first measurement of Ly$\alpha$ absorption from a young mini Neptune, HD 63433c \citep{zhang_2021}.  This planet is one of two transiting mini Neptunes orbiting a 400 Myr solar analogue.  Our non-detection of similar absorption from HD 63433b suggests that the inner planet may not possess a hydrogen and helium-rich envelope at all, in agreement with its shorter predicted atmospheric lifetime.  HD 63433 is a particularly favorable target for Ly$\alpha$ observations, as it is an active nearby star that is moving towards us.  This results in an unusually high flux in the near blue wing of the Ly$\alpha$ line, which is expected to contain most of the absorption signal from the planet's escaping atmosphere.  To date, Ly$\alpha$ absorption has only been conclusively detected for four planets aside from HD 63433c \citep{vidal-madjar_2008,bourrier_2013,lavie_2017,bourrier_2018b}, all of which are larger than Neptune.

The metastable helium triplet at 1083 nm can also be used to probe atmospheric mass loss \citep{oklopcic_2018,spake_2018}, and has the advantage of being readily accessible to ground-based observatories.  The strength of the absorption in this line is less than that in Ly$\alpha$, at most several percent, but this precision is achievable with many infrared spectrographs.  However, observations in this triplet are largely restricted to planets orbiting active K stars, which have the optimal UV spectral shape to produce a significant population of metastable helium \citep{oklopcic_2019}.  To date there have been multiple detections of outflows from close-in gas giant planets using this line, including the hot Jupiter HD 189733b \citep{salz_2018}, the inflated Saturn WASP-107b \citep{spake_2018,allart_2019}, and the warm Neptune GJ 3470b \citep{palle_2020}.  However, helium mass loss has never been securely detected for planets smaller than 4 $R_\Earth$ despite many attempts (i.e. \citealt{kasper_2020,gaidos_2020a,gaidos_2020b,zhang_2020,zhang_2021}).  Most of these published non-detections are of planets around old and inactive stars (HD 97658b, 55 Cnc e, GJ 1214, GJ 9857d), and the few observations of young planets were not very sensitive (K2-100b, K2-25b).  We also did not detect helium absorption from HD 63433c, despite the Ly$\alpha$ detection.  We suspect that the unfavorable host star type (G5) and underestimated outflow confinement mechanisms combined to suppress the size of the expected absorption signal in this line, while the unexpectedly high stellar variability in the 1083 nm line decreased the sensitivity of our observations.

The Transiting Exoplanet Survey Satellite (TESS) has vastly increased the sample of small transiting planets that are amenable to atmospheric characterization, including the aforementioned HD 63433 system.  TESS recently identified TOI 560.01 (HD 73583b), a $2.83 \pm 0.10$ $R_\Earth$, $10.1_{-3.0}^{+3.2} M_\Earth$ planet orbiting a K4V star with a 6.4 d period \citep{barragan_2021}. This star is young: the SuperWASP project's photometry reveals a robustly detected rotation period of 12 days, corresponding to a gyrochronological age of $\sim$600 Myr.  It is also close by, with a distance of 31.6 pc and a J band magnitude of 7.6.  The age and spectral type of the star, the size of the planet, and the closeness of the system combine to make TOI 560.01 an exceptionally favorable mini Neptune for probing helium mass loss. In addition, the planet has an outer companion at P=18.9 d with a very similar radius, making comparative mass loss studies possible.

In this paper, we use Keck/NIRSPEC to measure the helium absorption signal from TOI 560.01 and XMM to measure the star's high energy spectrum.  We then compare our helium measurement to predictions from 1D and 3D photoevaporative mass loss models.  We describe the observations in \S\ref{sec:data}, our analysis of the data in \S\ref{sec:analysis}, our reconstruction of stellar properties in \S\ref{sec:understanding_star}, and our modelling of the outflow in \S\ref{sec:modeling}.  We discuss in \S\ref{sec:discussion} and summarize our conclusions in \S\ref{sec:conclusion}.

\section{Observations and Data Reduction}
\label{sec:data}
\subsection{Keck/NIRSPEC}
\label{subsec:keck_reduction}
We used Keck/NIRSPEC to observe a full transit of TOI 560.01 on March 18, 2021 UTC and a partial transit on April 19, 2021 UTC (40\% of $T_{14}$).  All observations were obtained in Y band using the high resolution mode with the 12 x 0.432$\arcsec$ slit, resulting in a spectral resolution of 25,000.  We used 60 second exposure times, and adopted an ABBA nod pattern to help subtract background.  Because we only used one coadd per exposure, we achieved a high efficiency of 77\% for these observations.  On March 17, the sky was clear and the seeing was 0.6--0.7''.  We observed 0.7 h of pre-transit baseline, the 2.1 h transit, and 0.6 h of post-transit baseline.  We achieved a typical SNR of 170 per spectral pixel in one 60 second exposure.  On April 18, the seeing was better (0.5'' at both beginning and end of night), but there were sporadic cirrus clouds which caused large fluctuations in the water column density.  We observed 1.9 h of pre-transit baseline and 0.8 h of transit before the target sank too low to observe.  On this second night, we achieved a higher average SNR of 200 per spectral pixel, but with more weather-induced variability.  

We calibrated the raw images and extracted 1D spectra for each order using a custom Python pipeline designed for the upgraded NIRSPEC.  The pipeline is described in detail in \cite{zhang_2020}, but we summarize it here, along with the target-specific differences in our reduction for TOI 560.  First, the pipeline subtracts crosstalk from each raw frame.  Then, it calibrates the raw frames by computing a master flat, identifying bad pixels in the process.   It uses the master flat to compute a calibrated A-B difference image for each A/B pair.  The spectral trace containing the 1083 nm line (spanning 10,803--11,008 \ang) is identified, and we perform optimal spectral extraction to obtain spectra along with their errors.  A template is computed from a model stellar spectrum and model tellurics, with the stellar spectrum shifted in wavelength to account for the star's average Earth-relative radial velocity during that night.  We use the template to derive the wavelength solution for each individual spectrum.  

After extracting the 1D spectra, we remove tellurics.  In \cite{zhang_2020}, we did this by running SYSREM, which can be thought of as Principal Component Analysis with error bars \citep{mazeh_2007}.  However, we have subsequently found that while SYSREM performs excellently in removing tellurics, it also removes half the planetary signal unless the planet, like 55 Cnc e, undergoes extreme radial acceleration during the observation \citep{zhang_2021}.  Furthermore, due to an unfortunate coincidence, a strong telluric water line falls right on top of the helium line during both nights of our observations.  We therefore opted for a more conservative and physically motivated telluric removal method.  We used \texttt{molecfit} \citep{smette_2015}, which models tellurics using a meteorological model for Earth's temperature-pressure profile at the time and place of observation, but allows the user to fit for the rapidly varying water column density.  We fit for the water column density using narrow wavelength ranges containing strong telluric lines while avoiding any stellar lines.  We also fit for the continuum with a line.  The line spread profile is fixed to a Gaussian with a FWHM of 3.5 pixels, a value we settled upon after several fits to different spectra (with the FWHM as a free parameter) converged upon similar values.  3.5 pixels is close to the theoretical value of 3 pixels for the slit we used.  After fitting the water lines, \texttt{molecfit} produces a telluric corrected spectrum across the full wavelength range of the order containing the helium line, which we adopt for the rest of the analysis.

Having obtained wavelength calibrated and telluric corrected spectra, we interpolate all spectra onto a common wavelength grid with a uniformly logarithmic spacing of $\lambda/110,000$ and a range of 10,810--10,850 \ang.  We remove fringing by applying a notch filter twice, using exactly the same parameters described in \cite{zhang_2020}.  We divide each spectrum by the continuum, take the logarithm of the entire spectral grid, and subtract the mean of every row and column from that row and column.  This results in a residuals grid: a $N_{obs} \times N_{wav}$ grid of numbers representing the relative deviation of a pixel from the mean for that row and column.  For every column (wavelength), we subtract the mean of the out-of-transit part of the residuals image for that column; we then invert the residuals image.  The residuals image now gives excess absorption relative to the out-of-transit baseline.  However, there are continuum variations that cause structure in this image.  We mask out the helium line, mask out the strong lines because optimal extraction performs poorly with them \cite{zhang_2020}, fit a 3rd order polynomial to each row (epoch) with respect to wavelength, and subtract off the polynomial.

\subsection{XMM-Newton}
On April 21, XMM-Newton measured the star's X-ray and MUV spectrum, which are crucial for modeling photoevaporative mass loss and predicting the metastable helium population.  We observed the system for a total of 13 ks as part of XMM prop. ID 088287 (PI: Michael Zhang).  We configured the EPIC cameras to observe with the medium filter and small window, giving us 97\% observing efficiency on the two MOS CCDs and 71\% efficiency on the one pn CCD.  We configured the Optical Monitor to observe the star in the UVM2 filter ($231 \pm 48$ nm) for 7.2 ks and the UVW2 filter ($212 \pm 50$ nm) for 4.4 ks.  Although these observations are not simultaneous with the helium mass loss observations, they were taken a month after the first helium observation, a short time compared to typical stellar activity cycles.

To analyze XMM-Newton data, we download the raw Observation Data File (ODF) and use the Science Analysis System (SAS) provided by the XMM-Newton team to reduce it.  We run \texttt{xmmextractor}, to obtain spectra from the ODF with default settings.  For the Optical Monitor, SAS reports the count rate in the UVW2 and UVM2 filters, the two mid ultraviolet filters we selected.  For each of the three EPIC detectors, SAS generates the light curve and the background-subtracted spectrum.

For the pn detector, we find that the automatic reduction by \texttt{xmmextractor} leads to significantly negative fluxes at 0.23--0.28 keV, which is unphysical.  Therefore, we use SAS to manually reduce the pn data by defining the source region as a circle 17.5\arcsec in radius, and the background region as an annulus centered on the source with an inner radius of 20\arcsec and an outer radius of 35\arcsec.  Defined in this way, the spectrum no longer has significantly negative fluxes around 0.25 keV.

\section{Analysis of Helium Transit Observations}
\label{sec:analysis}

\begin{figure*}
    \subfigure {\includegraphics[width=0.5\textwidth]{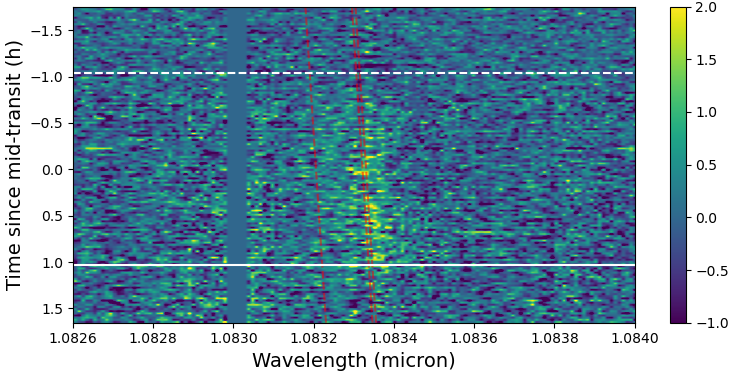}}\subfigure {\includegraphics[width=0.5\textwidth]{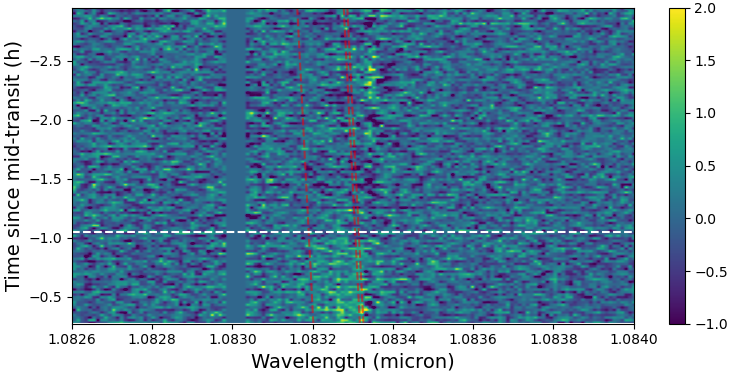}}
    \caption{Percent excess absorption from TOI 560.01 as a function of time and wavelength, for the first (left panel) and second (right panel) nights of observation.  The dashed white line indicates the beginning of the white light transit, while the solid white line indicates the end.  The red lines show the wavelengths of planetary helium absorption.  At 10,830 \ang is a deep stellar Si I line, which, like other strong lines, we mask as part of our analysis because optimal extraction deals poorly with very strong lines \citep{zhang_2020}.}
    \label{fig:excess_2D}
\end{figure*}

\begin{figure}
  \centering 
  \includegraphics
    [width=0.5\textwidth]{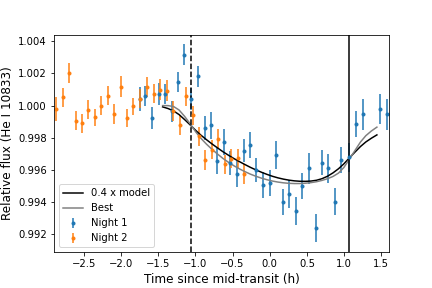}
    \caption{Light curve of the helium line (top) in a 1.5\ang bandpass centered on 10833.27\ang.  The dashed black line marks the beginning of the white light ingress, while the solid black line marks the end of white light egress.  We overplot the predicted light curves from the fiducial and best fit Microthena models as solid black and grey lines, respectively.  We rescale the amplitude of the fiducial model light curve by a factor of 0.4 to match the amplitude of the observed signal.}
\label{fig:helium_lc}
\end{figure}

\begin{figure}
  \centering 
  \includegraphics
    [width=0.5\textwidth]{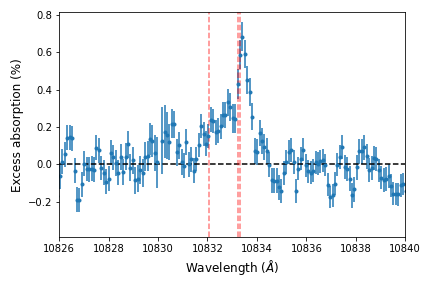}
    \caption{Excess absorption spectrum from the full transit observation on night 1, plotted in the planetary frame.}
\label{fig:excess_night1}
\end{figure}

\begin{figure}
  \centering 
  \includegraphics
    [width=0.5\textwidth]{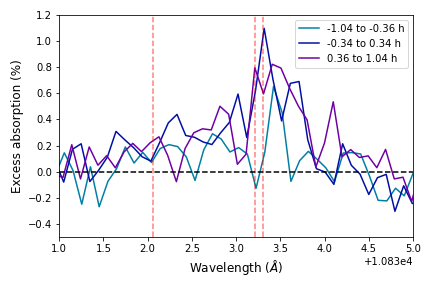}
    \caption{Excess absorption spectrum for each third of the transit on night 1. Wavelengths are in the planetary frame.}
\label{fig:three_segments_absorb}
\end{figure}

\begin{figure}
  \centering 
  \includegraphics
    [width=0.5\textwidth]{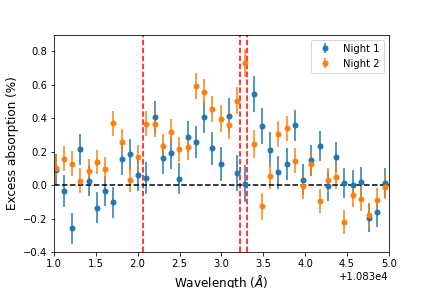}
    \caption{Comparison of the excess absorption spectrum on both nights, averaged over only the in-transit phases observed on night 2 (first 40\% of transit).}
\label{fig:excess_absorption_comp}
\end{figure}

Figure \ref{fig:excess_2D} shows the excess absorption in the 1083 nm metstable helium triplet as a function of time and wavelength, for both nights of observation with Keck/NIRSPEC.  To compute the excess absorption, we adopt a baseline that includes all pre-ingress spectra and (for the first night) all spectra taken more than 12 minutes after egress.  Ingress and egress are computed purely based on the white light ephemeris.  Figure \ref{fig:helium_lc} shows the band-integrated light curve for each night in the helium line.  We detect strong absorption starting from the white light ingress, reaching a maximum of 1.7\% half an hour after the midpoint of the white light transit, and declining quickly after white light egress.
The excess helium absorption begins on time on both nights, but ends $\sim$10 min late on the first night, the only night we captured the egress.  The band-integrated excess absorption is $\sim$0.7\%, and the equivalent width of the planetary absorption is $7 \pm 0.4$ m\AA{}.  The absorption spans at least 1.5\ang, corresponding to a velocity spread of 40 km/s--far higher than the escape velocity from the planetary surface of 15 km/s, or the escape velocity from the approximate helium absorption radius ($\sim4 R_p$) of 7.5 km/s.  This shows that the atmosphere is escaping, because velocity dispersion is the only significant broadening mechanism: natural broadening is of order 0.01 km/s, the equatorial rotational velocity of the planet is 0.2 km/s, and pressure broadening is $\sim$10$^{-11}$ km/s at picobar pressures because it is $\sim$10 km/s at atmospheric pressure \citep{niermann_2010}.  For unknown reasons, the helium flux rises by 0.2\% right before ingress on the first night.  Since we do not see the same brightening on the second night, and since there is no plausible way the planetary outflow can cause the star to brighten before the transit, we attribute this brightening to stellar activity.  The helium line is well known to be a tracer of chromospheric activity, which is more intense in younger stars.  In previous observations we monitored the young G5 star HD 63433 for two nights, and saw $>\sim0.2\%$ variability on both nights \citep{zhang_2021}.

In Figure \ref{fig:excess_night1}, we examine the wavelength-dependence of the transit-averaged excess absorption signal from the first night.  We find that it is redshifted by $\sim0.14$ \AA{} relative to the radial velocity of the planet, corresponding to a velocity of $\sim$4 km/s.  The 1D absorption spectrum peaks at $0.68 \pm 0.08$\% and declines quickly in the red wing but slowly in the blue wing, consistent with gas being pushed towards the observer by radiation pressure or stellar wind.  This extended tail of blue-shifted absorption is similar to that observed for WASP-107b \citep{allart_2019}, a planet with a far more extended egress in the integrated helium light curve.  In Figure \ref{fig:three_segments_absorb}, we divide the transit into thirds and show the excess absorption spectrum for each third.  Consistent with the other plots, there is slightly redshifted absorption in each part, with the final third having much stronger absorption than the first third.

In Figure \ref{fig:excess_absorption_comp}, we compare the excess absorption spectrum for the two nights averaged over the in-transit phases observed on the second night.  We find that the absorption observed on the second night appears slightly stronger and more blueshifted than it did during the equivalent time window on the first night.  The two nights are otherwise consistent.  To quantify the significance of the differences between the two nights, we used nested sampling as implemented by \texttt{dynesty} \citep{speagle_2019} to fit a Gaussian to the excess absorption spectra plotted in Figure \ref{fig:excess_absorption_comp}, with 3 free parameters: amplitude, standard deviation, and mean.  We find a mean position of $10833.06 \pm 0.15$ \ang on night 1 and $10,832.70 \pm 0.06$ \ang on night 2, a difference of $0.36 \pm 0.16$ \ang ($10 \pm 4$ km/s in velocity space).  The amplitude was $0.22 \pm 0.04\%$ for night 1 and $0.40 \pm 0.04\%$ on night 2, for a difference of $0.18 \pm 0.06\%$.  These statistical tests confirm what visual inspection shows: on night 2, the helium absorption was marginally stronger and marginally bluer.  Further observations are necessary to determine whether this variability is due to underestimated error bars, a consequence of poorly understood stellar variability, or a change in the properties of TOI 560.01's outflow.

On the second night of observation, the stellar helium line was slightly narrower (by 8\%) than on the first night, but of indistinguishable depth (to within 1.5\%) and position (to within 0.5 km/s).  Minor differences in the stellar line do not affect our results for the planetary excess absorption, which is computed by comparing in-transit and out-of-transit spectra on the same night.

\subsection{Possible causes of the observed redshift}
The redshifted absorption peak seen during the first night (Figure \ref{fig:excess_2D} and \ref{fig:excess_night1}) is unusual.  Aside from HAT-P-32b \citep{czesla_2021}, none of the gas giant planets with spectrally resolved helium observations have exhibited a similar redshift \citep[e.g.,][]{salz_2018,allart_2019,palle_2020}, and material flowing away from the star should appear blueshifted during transit.  If the observed redshift is due to the geometry of TOI 560.01's atmospheric outflow, it would suggest that TOI 560.01 has unusual outflow properties, perhaps implying an unexpected weak stellar wind.  However, we must first ascertain whether the apparent redshift is real.

The apparent redshift is unlikely to be due to wavelength calibration uncertainties.  NIRSPEC does experience wavelength drift over the course of a night \citep{kasper_2020}, but our wavelength calibration is performed independently on every spectrum, so drift does not affect us.  In addition, we computed the wavelength solutions with an alternate method that uses the median observed spectrum instead of the theoretical spectrum as a template.  Comparing the results, we conclude that the wavelength solution for each spectrum is accurate to at least 1 km/s.

The apparent redshift is also not due to ephemeris uncertainties.  The ephemeris predicts the transit mid-point on the first night with a 1$\sigma$ uncertainty of 1 minute, during which time the planet accelerates by 0.07 km/s.  This is much smaller than the observed redshift of $\sim$4 km/s.  Similarly, uncertainties on the stellar mass and semimajor axis can only change the acceleration of the planet by $\sim7\%$.  Since the planet accelerates from $-4.5$ km/s to $+4.5$ km/s during the transit, a 7\% change is insufficient to account for the observed redshift.

The final source of uncertainty is the eccentricity.  To obtain the radial velocity of the planet, we assumed a perfectly circular orbit.  However, \cite{mills_2019} used the transit durations of 1000 Kepler planets, combined with accurate stellar radii from the California-Kepler Survey and Gaia, to statistically infer typical eccentricities.  They obtained a mean eccentricity of 0.05 for systems with multiple transiting planets.  The star-planet distance changes at a maximum rate of $K_p e$ where $K_p$ is the orbital speed; since $K_p$ = 102 km/s for TOI 560.01, an eccentricity of 0.05 can cause an apparent helium signal redshift of up to 5.1 km/s.  \cite{barragan_2021} constrained the eccentricity to $0.10_{-0.07}^{+0.08}$ using a joint fit to the radial velocity and light curve, which is consistent with both 0 and 0.05.  However, their radial velocity fit results in a planetary radial velocity of $-7.4_{-7.0}^{+9.3}$ km/s at mid-transit, which is in the wrong direction, albeit consistent with a 4 km/s redshift to within 2$\sigma$.  We are therefore unable to differentiate between a redshift caused by a non-zero orbital eccentricity and one caused by the geometry of the outflow.

\subsection{Telluric absorption}

\begin{figure}
    \includegraphics[width=0.5\textwidth]{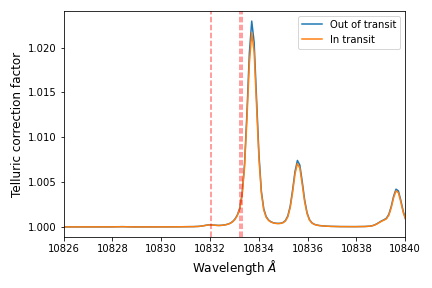}
    \caption{Median telluric correction factors inferred by molecfit, for the in-transit and out-of-transit spectra.  The wavelengths of the helium triplet are indicated with vertical red lines.}
    \label{fig:telluric_correction_factors}
\end{figure}

\begin{figure}
    \subfigure {\includegraphics[width=0.5\textwidth]{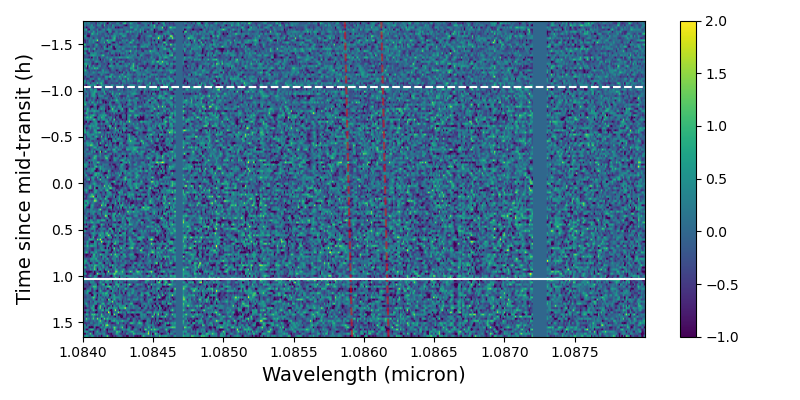}}
    \subfigure {\includegraphics[width=0.47\textwidth]{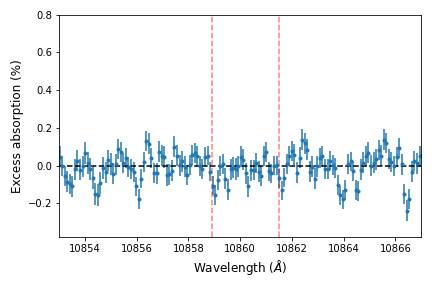}}
    \caption{To explore the effect of telluric correction on the helium signal, which overlaps with a telluric line, we choose two other telluric lines: one slightly weaker, and one significantly stronger.  We show the 2D excess absorption spectrum in the stellar frame (top) and the 1D excess absorption spectrum in the planetary frame (bottom), with the wavelengths of the telluric lines marked in red; the bluer line is the weaker.  These plots are analogous to Figure \ref{fig:excess_2D} and Figure \ref{fig:excess_night1}.}
    \label{fig:excess_telluric}
\end{figure}

As we mention in Subsection \ref{subsec:keck_reduction}, a telluric water absorption line overlaps with the red end of the stellar helium line, and we use molecfit to correct for this telluric absorption.  One might wonder whether imperfect telluric correction impacts our analysis, and in particular, whether it might be the cause of the apparent redshift in the absorption signal.

Figure \ref{fig:telluric_correction_factors} shows the median correction that molecfit applied to the in-transit and out-of-transit spectra.  Although the total correction was, at maximum, 2\%--which is comparable to the highest excess absorption from the planet--the correction is nearly identical for the in-transit  and out-of-transit spectra, with a difference of only 0.12\%.  However, the out-of-transit spectra span a larger range of airmass than the in-transit spectra, so comparing only the median correction factors does not tell the whole story.

To get a better idea of the effects of telluric correction, we looked at two other telluric lines redward of the helium line: one with a slightly lower absorption depth (1.5\%), and one with a significantly higher absorption depth (4\%).  Figure \ref{fig:excess_telluric} shows the excess absorption spectrum (both 2D and 1D) in the region of wavelength space around these lines.  For both lines, there are no statistically significant features in either plot.  In the 1D spectrum, there does appear to be small (0.1\%) dips redward of both telluric lines.  These dips are probably coincidental, but if not, they would only strengthen our conclusion that the helium absorption is redshifted.

\subsection{RM and CLV}
The Rossiter-Mclaughlin (RM) effect and center-to-limb variations (CLV) plague high-resolution transit spectroscopy of giant exoplanets.  They introduce pseudo-signals which can be a substantial portion of the helium absorption signal for planets like HD 189733b \citep{salz_2018}.

The RM effect occurs when the planet or its escaping atmosphere blocks out a portion of the rotating stellar disk.  Because the portion it blocks out has a non-zero rotational velocity, the star appears to experience a radial velocity change during the course of the transit.  Combining the radius and rotational period of TOI 560, we calculate a rotational speed of 2.8 km/s.  If escaping atmosphere blocks 0.7\% of the limb, an apparent radial velocity shift of (0.7\%)(2.8) = 0.02 km/s is created.   To estimate the effect this has on the excess absorption, we multiply by the maximum derivative of the relative flux in the vicinity of the helium line, $\frac{dln(F)}{dv} = 0.02$ km$^{-1}$ s.  We obtain 0.04\%, well below the average excess absorption of 0.7\%, and below our noise.

The CLV effect occurs because line centers experience different limb darkening from the continuum.  Standard stellar spectra libraries such as PHOENIX \citep{husser_2013} and MARCS \citep{van_eck_2017} do not model the chromosphere, and therefore do not include the helium line.  Without knowing how the helium line depth varies across the stellar surface, it is impossible to precisely model the CLV, but we can do an approximate calculation.  \cite{de_jager_1966} reported that for the Sun, the center ($\theta=0^\circ$) has a 10833\ang line depth of 6\% while the edge ($\theta=70^\circ$) has a line depth of 10\%.  TOI 560 has a much deeper 10833\ang line, at 30\%.  If the center of the star has a line depth of 20\% and the edge has a depth of 40\%, when the escaping atmosphere blocks 1.7\% of the center of the stellar disk, the stellar pseudo-signal would increase the apparent helium signal by (10\%)(1.7\%) = 0.17\%.  This is not negligible, but it is a small fraction of the total signal.  It is not possible under any circumstances for the entire helium signal to be due to CLV because the white light transit depth is only 0.16\%, far smaller than the measured helium signal.

\section{The star}
\label{sec:understanding_star}
\subsection{Age}
We estimate the star's age with gyrochronology.  The star has a rotation period of $12.2 \pm 0.3$ d and a mass of $0.71 \pm 0.02 M_\odot$ \citep{barragan_2021}.  It has a B-V color of $1.112 \pm 0.001$, from the Hipparcos input catalog \citep{turon_1993}.  Using \cite{schlaufman_2010}, which relates age to stellar mass and rotation period, we obtain an age of $635 \pm 40$ Myr.  Using \cite{mamajek_2008}, which relates age to B-V color and rotation period, we obtain $540 \pm 80$ Myr.  The error bars for the first estimate are purely statistical, with no model uncertainty, while the error bars for the second estimate include the statistical uncertainties of the model parameters.

\subsection{X-ray observations}
\begin{figure}
  \centering 
  \subfigure {\includegraphics
    [width=0.5\textwidth]{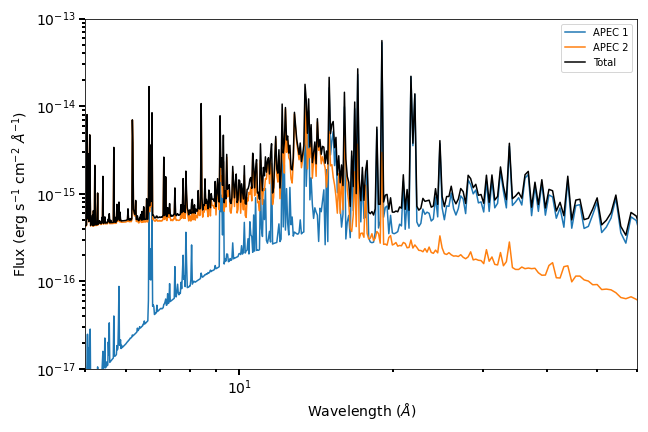}}
   \subfigure {\includegraphics
    [width=0.5\textwidth]{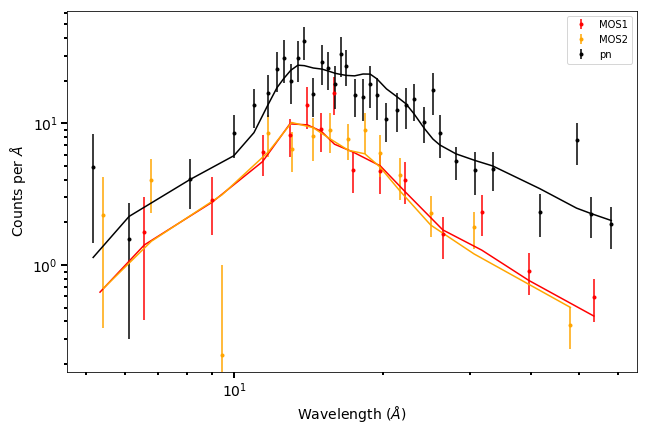}}
    \caption{Top: Best fit model spectrum from our XMM EPIC observations of TOI 560.  The total spectrum as observed from Earth (black) is the sum of a low-temperature component (APEC 1) and a high-temperature component (APEC 2).  Bottom: the data and the folded models, which take into account the instrumental throughput and line spread function.}
\label{fig:xmm_epic}
\end{figure}

We analyzed the XMM data using the approach described in \cite{zhang_2021}, which we summarize here.  We used \texttt{xspec} to fit a model consisting of two components of optically thin, collisional plasma in equilibrium.  We also tried one and three components, but found that two components minimizes the Bayesian Information Criterion.  We show the data and fitted model in Figure \ref{fig:xmm_epic}, and list the corresponding model parameters in Table \ref{table:xmm_params}.  The model is a good fit to the data for all three EPIC detectors, and the fit parameters are consistent with expectations for a moderately active star.  As discussed in \cite{wood_2010} and \cite{zhang_2021}, coronal metallicities are often lower than the equivalent photospheric metallicities.

\begin{table}[ht]
  \centering
  \caption{Model parameters for XMM-Newton data}
  \begin{tabular}{c C}
  \hline
  	  Parameter & \text{Value}\\
      \hline
      Metallicity & 0.60 \pm 0.26\\
      kT$_1$ (keV) & 0.25 \pm 0.03\\
      EM$_1$ (cm$^{-3}$) & 7.3 \pm 2.5 \times 10^{50}\\
      kT$_2$ (keV) & 0.94 \pm 0.13\\
      EM$_2$ (cm$^{-3}$) & 2.2 \pm 0.7 \times 10^{50}\\
      Flux$^*$ (erg/s/cm$^2$) & 1.05 \pm 0.06 \times 10^{-13}\\
      \hline
  \end{tabular}
  \tablecomments{$^*$Derived, not a fit parameter.  For the range 5--100 \AA{} (0.124--2.48 keV).}
  \label{table:xmm_params}
\end{table}

\begin{figure}
    \includegraphics[width=0.5\textwidth]{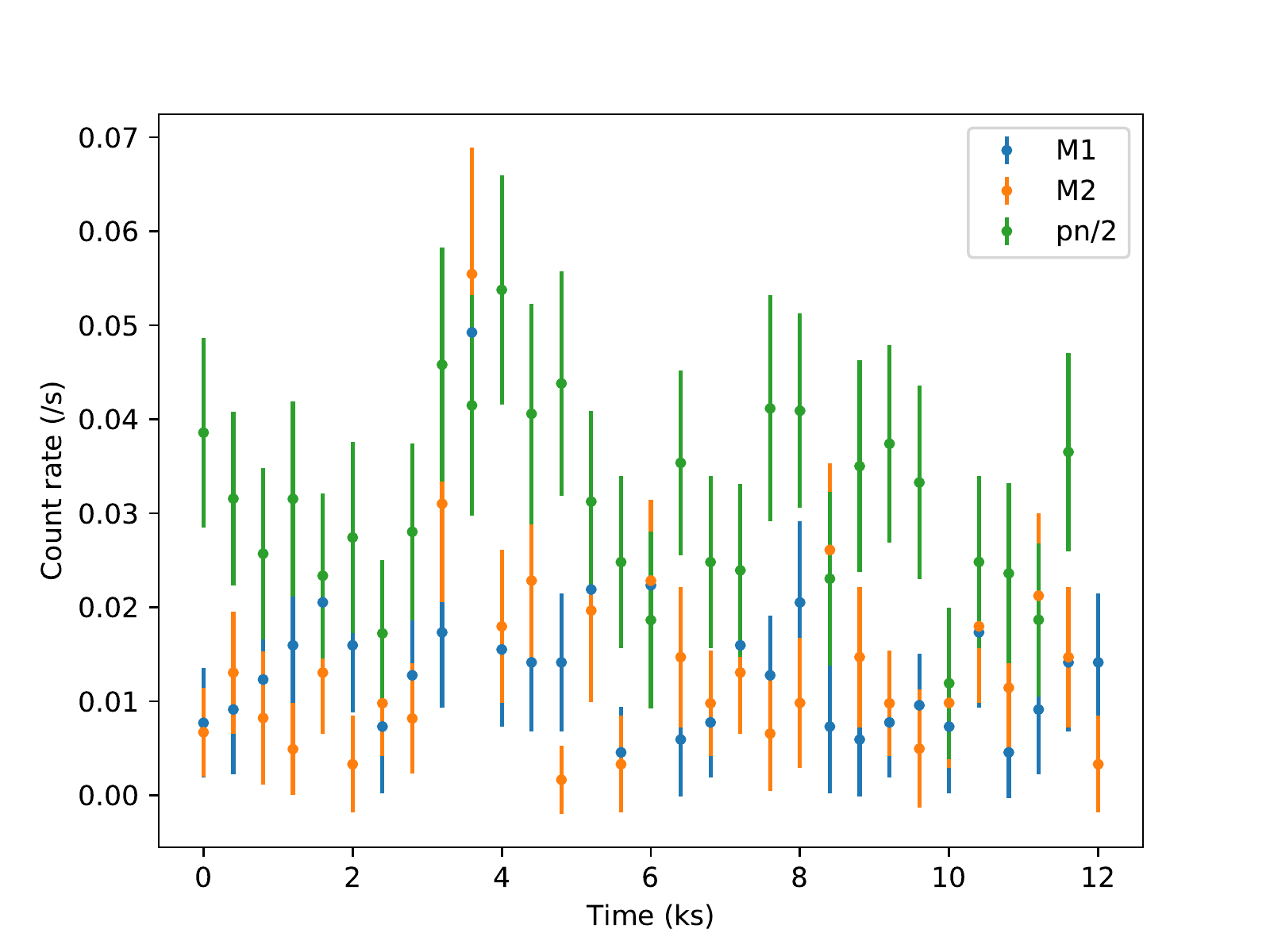}
    \caption{Background-subtracted X-ray light curves of TOI 560, recorded by the three EPIC cameras.}
    \label{fig:xmm_epic_lc}
\end{figure}

From the model fits, we derive a 5--100 \AA{} flux of $1.05 \pm 0.06 \times 10^{-13}$ erg/s/cm$^2$.  However, the star's intrinsic variability is likely to be much higher than the reported precision for this flux measurement.  Figure \ref{fig:xmm_epic_lc} shows the light curve measured by the EPIC detectors during the 12 ks observation.  As is typical for X-ray observations of active stars, the light curve is variable at the 20\% level on kilosecond time scales.

\subsection{Ly$\alpha$ and EUV}
The star's extreme UV flux (roughly 100--912 \AA{}) is the predominant driver of photoevaporative mass loss.  Unfortunately, no EUV telescopes currently exist, and strong interstellar absorption makes it difficult to accurately measure EUV flux from even the closest stars.  To obtain the EUV spectrum, we use the scaling relations of \cite{linsky_2014}, which give the EUV flux in 100 \AA{} bins with respect to the Ly$\alpha$ flux.  Unfortunately, there are also no measurements of the star's Ly$\alpha$ flux.  To compute the flux, we use \cite{linsky_2013}, which provides a scaling relation with respect to the rotation period and another with respect to the X-ray flux, with mean dispersions of 32\% and 22\%, respectively.  For TOI 560, both scaling relations give the same Ly$\alpha$ flux: 18 erg s$^{-1}$ cm$^{-2}$ at 1 AU.

\subsection{MUV}
The 1230--2588 \AA{} stellar flux (which we will call ``MUV'') ionizes triplet state helium, but does not contribute to producing it because it cannot ionize hydrogen or helium.  It therefore plays a crucial role in determining the size of the metastable helium population.  To quantify the MUV luminosity, we observed the star in two MUV filters using XMM-Newton's Optical Monitor.  The Optical Monitor recorded a count rate of $0.654 \pm 0.011$ s$^{-1}$ in the UVM2 filter and $0.565 \pm 0.015$ s$^{-1}$ in the UVW2 filter.  Correcting for the minor coincidence losses, we obtain count rates of $0.679 \pm 0.011$ s$^{-1}$ and $0.585 \pm 0.015$ s$^{-1}$, respectively.

As a starting point for the UV spectrum, we used a PHOENIX model \citep{husser_2013} for a star with $T_{\rm eff}=4600$ K, log(g)=4.5, and [M/H]=0.  We then rescaled this model to reflect the radius of the star, its distance, and the effective areas of the filters as a function of wavelength, and used it to predict the count rate in each filter.  We obtain $0.41 s^{-1}$ for UVM2 and $0.45 s^{-1}$ for UVW2, substantially below the measured values.  To rectify the mismatch, we multiplied the PHOENIX UV flux (1230--3500 \AA{}) by a factor of 1.67, bringing the predicted count rates (0.677 and 0.586 s$^{-1}$) into exact alignment with our measurements.

\subsection{Reconstructed stellar spectrum}
\label{subsec:stellar_spec}

\begin{figure}
    \includegraphics[width=0.5\textwidth]{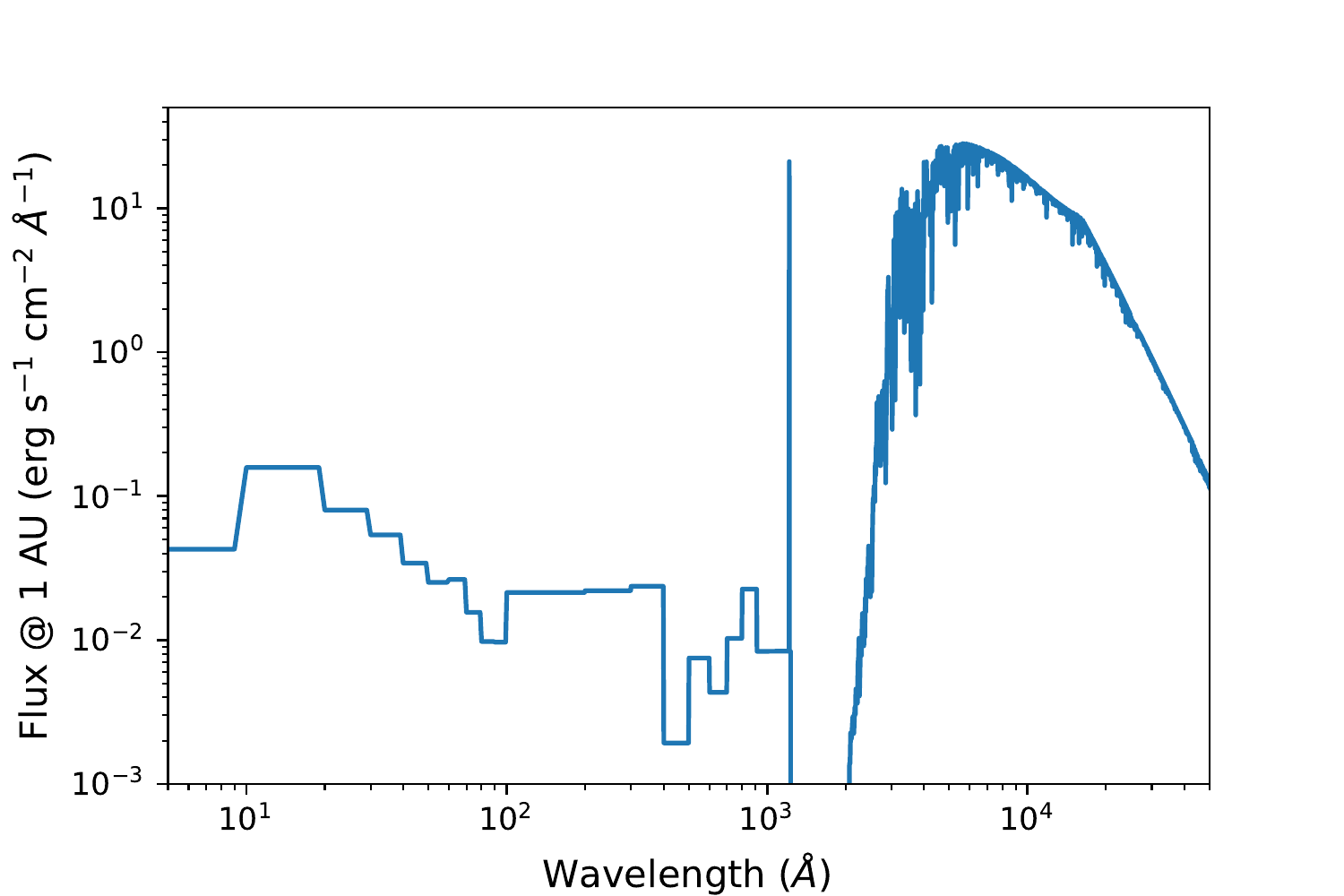}
    \caption{Fiducial stellar spectrum used in our mass loss modeling.  We bin the X-ray data for better visibility in this figure.  The errors associated with the spectrum are listed in Table \ref{table:band_fluxes}, and were computed using the methodology of \cite{zhang_2021}.}
    \label{fig:stellar_spectrum}
\end{figure}

\begin{table}[ht]
  \centering
  \caption{Band-integrated fluxes}
  \begin{tabular}{c C C}
  \hline
  	  Band & \text{Wavelengths (\AA)} & \text{Flux at 1 AU (cgs)}\\
      \hline
      X-ray & 5-100 & 4.4 \pm 2.2\\
      EUV & 100-912 & 14 \pm 4\\
      Ly$\alpha$ & 1214-1217 & 18 \pm 4\\
      MUV & 1230-2588 & 11 \pm 2\\
      Total & 5-50,000 & 2.47 \pm 0.13 \times 10^5\\
      \hline
  \end{tabular}
  \label{table:band_fluxes}
\end{table}

Figure \ref{fig:stellar_spectrum} shows the reconstructed stellar spectrum, while Table \ref{table:band_fluxes} lists the corresponding band-integrated fluxes.  The error bars on the band-integrated fluxes, which are very approximate, are computed using the methodology of \cite{zhang_2021}.  We find that the ratio of MUV to XUV (X-ray + EUV) flux for this star is 0.6, which is relatively low.  For the young solar analogue HD 63433, this ratio is 63 \citep{zhang_2021}.  As expected, K-type stars like TOI 560 have lower MUV-to-XUV ratios than G stars \citep{oklopcic_2019}.  Low MUV-to-XUV ratios are more favorable for helium observations because the MUV flux destroys triplet helium while the XUV flux, although capable of destroying triplet helium, also ionizes hydrogen and helium.  This creates the electrons and ionized helium which then recombine to produce triplet state helium.

\subsection{Stellar wind}\label{subsec:stellar_wind}
The stellar wind shapes the planetary outflow by confining it and pushing it away from the star, creating a Coriolis force that can lead to the formation of a comet-like tail.  However, the density and speed of the solar wind are highly variable with time, and there are few observational constraints on winds from other stars.  To infer the stellar wind conditions, we rely on \cite{wood_2005b}, which used astrospheric absorption to characterize the mass loss rates of a handful of nearby stars.  This absorption occurs where the stellar wind collides with the interstellar medium and charge exchange creates a population of hot neutral hydrogen, which then absorbs the stellar Ly$\alpha$ emission line.  In order to translate a measured Ly$\alpha$ profile into a mass loss rate, one must accurately model the intrinsic stellar Ly$\alpha$ emission, the population of energetic neutral atoms, and the neutral hydrogen of the interstellar medium along the line of sight, none of which are trivial problems.  Despite these challenges, \cite{wood_2005b} were able to use their measured mass loss rates to derive a scaling relation with the star's X-ray flux.  We use this relation to obtain an estimated mass loss rate of 11 times solar for TOI 560.  We further assume that the wind speed is comparable to that of the Sun (400 km/s), and use these values for our fiducial 3D models described in \S\ref{subsec:3d_models}.

The mass loss rate we derived is consistent with the new data presented by \cite{wood_2021} (their Figure 10).  However, the new data also shows that among the 7 GK dwarfs with an X-ray flux similar to TOI 560, the range in inferred mass loss rate per unit surface area is over 200x.  This means it is unfortunately impossible to predict the stellar mass loss rate to even an order of magnitude with any confidence.

\section{Modeling}
\label{sec:modeling}
We model the outflow using two different hydrodynamic codes: The PLUTO-CLOUDY Interface (TPCI), a 1D code developed by \cite{salz_2015}; and Microthena, a 3D code developed by \cite{wang_2018}.  Both codes have been extensively used to study photoevaporation (e.g. \citealt{salz_2015b,salz_2016,kasper_2020,zhang_2020} for TPCI; \citealt{wang_2020,wang_2021,zhang_2021} for Microthena).  As we discuss below, we view the 3D simulations as a more accurate predictor of the observed helium signal during transit, because photoevaporation is fundamentally a 3D problem.  Different areas of the planet receive different stellar irradiation levels, and the stellar wind cannot be ignored for young and active stars.  Nevertheless, we find it useful to run both models for two reasons.  First, 1D models much simpler than TPCI are widely used to interpret helium observations, most commonly variants of the Parker wind model proposed by \cite{oklopcic_2018} (e.g. \citealt{kasper_2020}).  By running both 1D and 3D models for TOI 560.01 and comparing their predictions, we can evaluate the magnitude of the error introduced by assumptions of radial symmetry for this planet.  Second, there are aspects of the problem that TPCI handles better than our 3D models, which must use a simplified physics framework in order to remain computationally tractable.  In particular, Microthena divides radiation into just seven energy bins and only computes atomic line cooling from a small number of species, while TPCI propagates the entire continuum and computes line cooling from all neutral and ionized species.    We therefore take advantage of TPCI's sophisticated radiative transfer to explore the qualitative effects of increasing metallicity (up to 100$\times$ solar) on the absorption signature.  This is a topic that has been relatively unexplored in the literature to date,  but TOI 560.01's small size and high bulk density mean that it could easily host a metal-rich atmosphere.

\subsection{1D TPCI models}
TPCI is a combination of two sophisticated and widely used codes: the hydrodynamic solver PLUTO \citep{mignone_2007}, and the plasma simulation and spectral synthesis code CLOUDY \citep{ferland_2013}.  CLOUDY computes the equation of state and heating and cooling rates and provides these to PLUTO; PLUTO evolves the fluid using the Euler equations and provides the new conditions to CLOUDY, restarting the cycle.

We set up the simulations following the methodology of \cite{zhang_2020}, which in turn is based on the methodology of \cite{salz_2015b}.  Briefly, we set the lower boundary condition at a radius of 2.9 $R_\Earth$, a particle number density of $10^{14}$ cm$^{-3}$, and a pressure computed from the number density and a pressure of $P = n k_B T_{eq}=10.2$ dynes/cm$^2$ (assuming $T_{eq}=740$ K).  The planetary mass is fixed at 9 $M_\Earth$, derived from the mass-radius relation of \cite{chen_2016} and consistent with the RV-measured mass of $10.06_{-2.98}^{+3.24} M_\Earth$.  The stellar spectrum is derived in Section \ref{sec:understanding_star}.  In \cite{zhang_2020}, we did not include any metals or molecules; here, we continue to neglect molecules, but include metals with a solar abundance greater than $10^{-5}$: C, N, O, Ne, Mg, Si, S, Fe.  For our four simulations, the abundances of these metals are scaled from solar by a factor of 1, 10, 30, and 100, and the mean molecular weight is adjusted accordingly.  At Z=1, hydrogen makes up 92\% of the atoms, helium 8\%, and the metals less than 0.1\%.  By weight, the ratios are 74\%, 25\%, and 1.3\%.  At Z=100, even though the metals comprise only 9\% of the atoms, they comprise 57\% of the mass; hydrogen and helium are only 32\% and 11\%, respectively.  At this metallicity, hydrogen and helium become minor species by mass.

We evolve each simulation for at least 100 time units without advection, where 1 unit is approximately the sound crossing time of $R_p/$(10 km/s).  We then evolve the simulation for at least 10 time units with advection turned on (100 units for Z=1, 10, and 30), at which point we perform a visual inspection to ensure convergence has been reached.  We use the conditions at the last timestep to compute the helium absorption depth.  Since the outflow is 3D and we only simulate the substellar point, we must rescale the model predictions to approximate the 3D geometry.  \cite{stone_2009} showed that the mass loss rate is higher by a factor of 4 in the spherically symmetric simulation, and we therefore adopt this as our rescaling factor, as was done by \cite{salz_2016}.  We implement the rescaling by dividing both the velocities and the densities by two, because this gives a helium line width comparable to the observed width.  Given that photoevaporation is inherently a 3D process, there is no ``right'' way to obtain a transit spectrum from a 1D simulation.

\begin{figure}
    \subfigure {\includegraphics[width=0.5\textwidth]{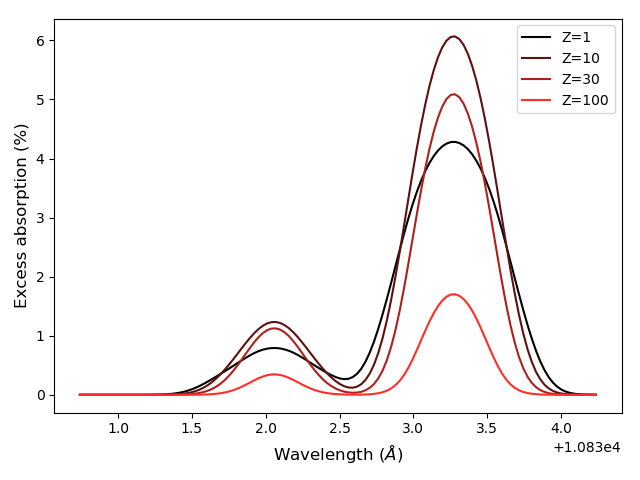}}
    \subfigure {\includegraphics[width=0.5\textwidth]{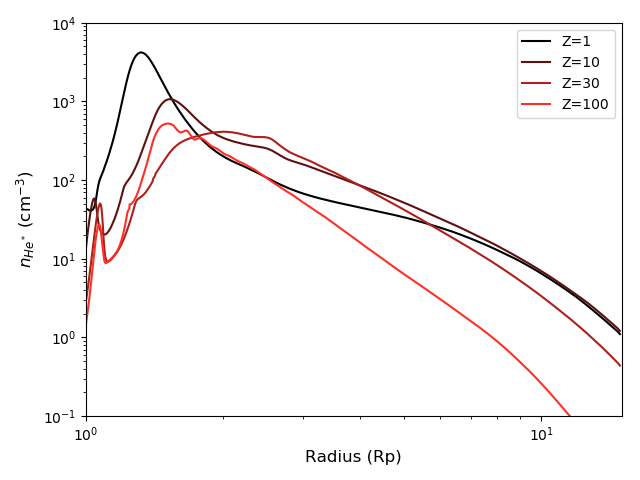}}
    \subfigure {\includegraphics[width=0.5\textwidth]{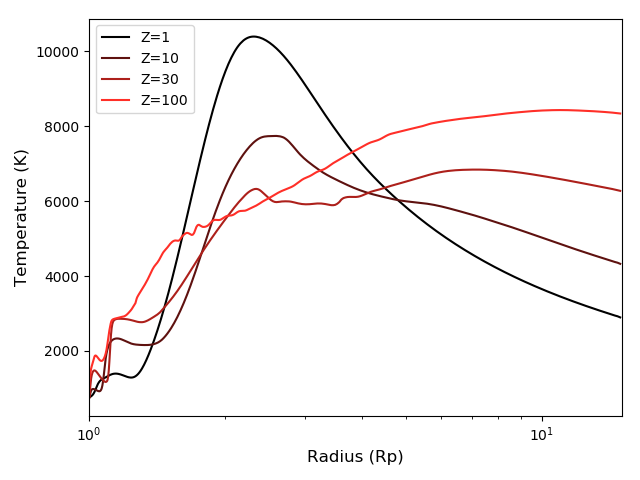}}
    \caption{Excess absorption spectrum (top), triplet helium number density (middle), and temperatures (bottom) predicted by TPCI as a function of metallicity.}
    \label{fig:excess_tpci}
\end{figure}

Figure \ref{fig:excess_tpci} shows how the the resulting excess absorption spectrum varies as a function of metallicity, while Table \ref{table:models_summary} list the mass loss rate, peak absorption, and equivalent width of the absorption.  Despite our rescaling to account for 3D effects, we find that our 1D models systematically overpredict the measured helium absorption signal, which peaks at 1.7\%.  The model with a metallicity of 100$\times$ solar has the weakest predicted helium absorption signal.  However, Figure \ref{fig:excess_tpci} shows that the predicted helium absorption signal does not decrease monotonically with increasing metallicity.  In fact, the helium signals for the 10$\times$ and 30$\times$ solar metallicity models are stronger than for the solar metallicity model.  The mass loss rate follows a similar pattern: we obtain 0.32, 0.42, 0.37, and 0.13 $M_\Earth$/Gyr for the 1$\times$, 10$\times$, 30$\times$, and 100$\times$ metallicities, respectively.  The velocity at the edge of our simulation (15 $R_p$) decreases with increasing metallicity, from 29 km/s to 20 km/s, while the location of the sonic point increases slightly with metallicity, with a minimum value of 2.5 $R_p$ at Z=1 and a maximum value of 3.1 $R_p$ at Z=100.  The density is fairly similar: 50\% higher for Z=10 and 30 than for Z=1, 40\% lower for Z=100 than for Z=1.

\begin{figure*}
    \subfigure {\includegraphics[width=0.5\textwidth]{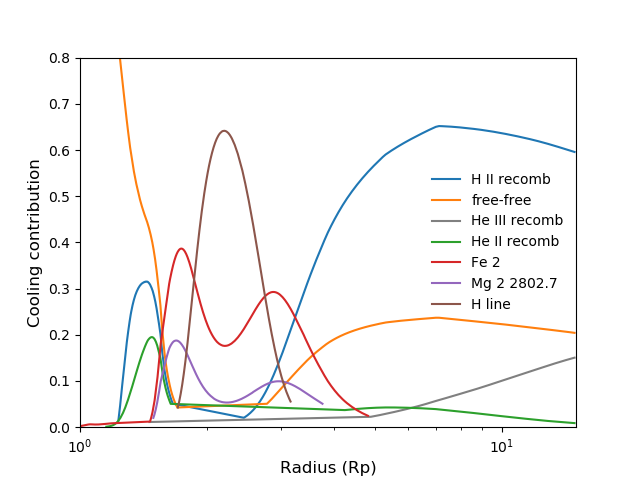}}\subfigure {\includegraphics[width=0.5\textwidth]{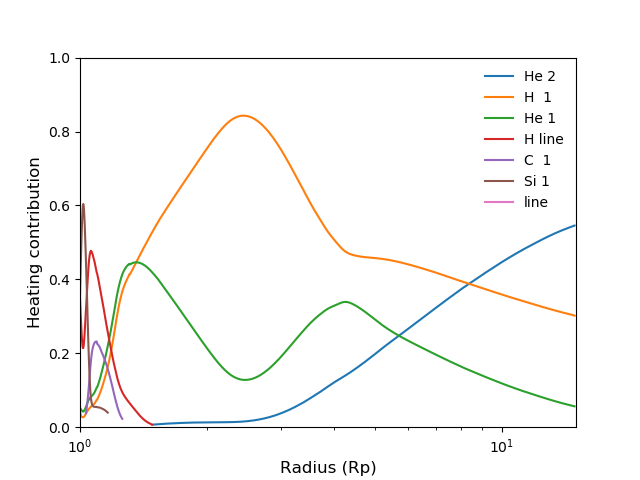}}
    \subfigure {\includegraphics[width=0.5\textwidth]{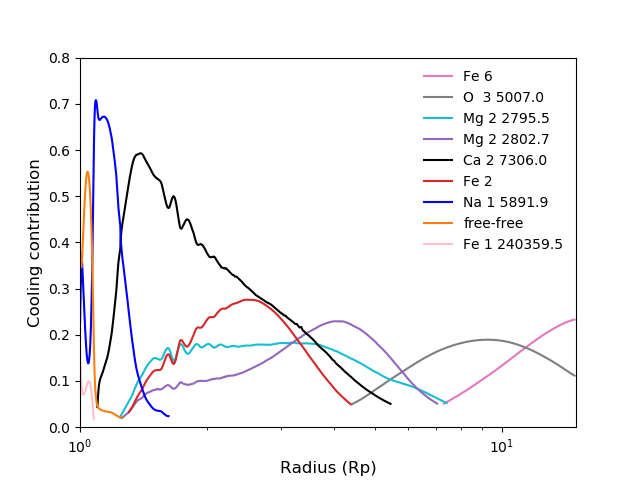}}\subfigure {\includegraphics[width=0.5\textwidth]{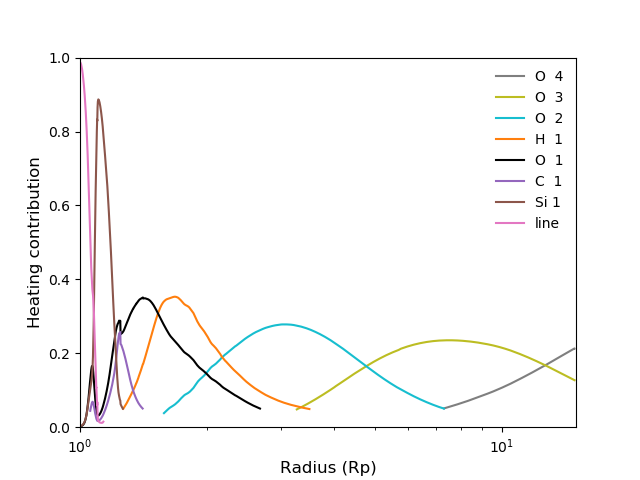}}
    \caption{Top row: radiative cooling (left panel) and radiative heating (right panel) contributions for our 1D solar metallicity TPCI model.  Bottom row: cooling (left panel) and heating (right panel) contributions for our 100$\times$ solar metallicity TPCI model.  The labels are as provided by TPCI.  For cooling, a label like O 3 5007.0 means the OIII line at 5007\ang.  For heating, a label like O 3 means photoionization of OIII, while ``line'' refers to line heating.  While hydrogen and helium are dominant in both heating and cooling at Z=1, a variety of metals in a variety of ionization states dominate heating and cooling at Z=100.}
    \label{fig:heating_cooling_contributions}
\end{figure*}

\begin{figure*}
    \subfigure {\includegraphics[width=0.5\textwidth]{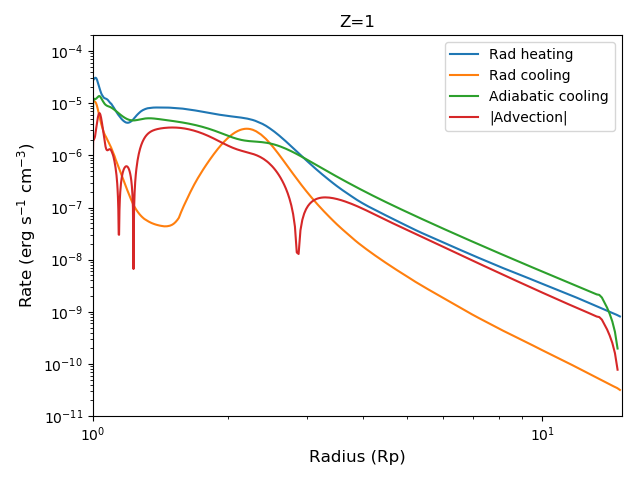}}\subfigure {\includegraphics[width=0.5\textwidth]{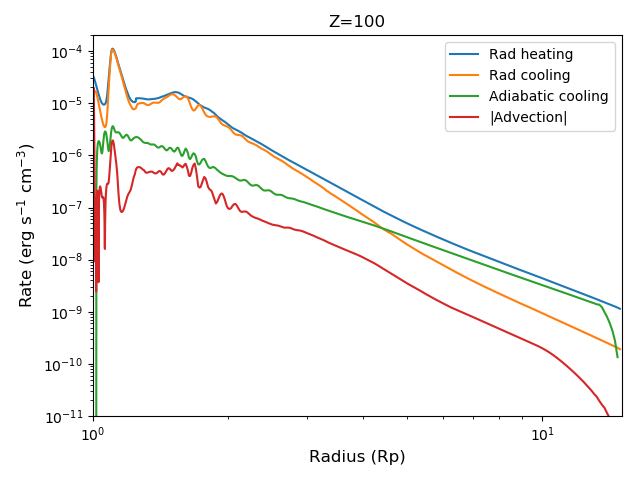}}
    \caption{Comparison of the four terms of the energy conservation equation (Equation \ref{eq:energy_conservation}).  The advection term switches sign, so we plot the absolute value.  The oscillations in the Z=100 plot are numerical artifacts.  The radiative heating and cooling terms are further dissected in Figure \ref{fig:heating_cooling_contributions}.}
    \label{fig:thermal_terms}
\end{figure*}

While the density in the far regions is similar between the 1$\times$ and 100$\times$ solar metallicity models, $n_e$ is 2.5$\times$ lower in the 100$\times$ solar model, while $n_{\rm HeII}$ is 6$\times$ lower.  Since triplet helium is primarily produced by recombination at a rate of $n_e n_{\rm HeII} \alpha$ and destroyed by photoionization, and the recombination coefficient $\alpha$ is a factor of two lower in the higher metallicity case due to the higher temperature \citep{pequignot_1991}, one would expect the triplet helium density in the outer regions to be 30$\times$ lower--in approximate agreement with the TPCI models.  We ascribe the low $n_e$ at high metallicity to the lower mass density, and to the lower number of easily ionizable electrons per unit mass.  For example, if we define easily ionizable as having an ionization energy smaller than 54 eV (the second ionization energy of helium), hydrogen and helium have 1 easily ionizable electron per amu, carbon has 0.25, neon has 0.1, and iron has 0.07.  The lower $n_{\rm HeII}$ at high metallicity is due to a combination of helium being less abundant (11\% by mass vs. 25\%) and the second-ionized fraction being higher (77\% vs 56\%), leaving fewer helium atoms in the singly ionized state.  The latter, in turn, is due to a combination of $n_e$ being 2.5$\times$ lower, and the temperature-dependent helium recombination coefficient being a factor of two lower.  In short, helium absorption drops dramatically at high metallicities largely because metals begin dominating hydrogen and helium by mass, but also due to changes in heating and cooling rates.

The temperature profile is governed by the equation of energy conservation \citep{murray-clay_2009}:

\begin{align}
\label{eq:energy_conservation}
    \rho v \frac{\partial}{\partial r} \Big[\frac{kT}{(\gamma - 1) \mu}\Big] = \frac{k_B T v}{\mu}\frac{\partial \rho}{\partial r} + \Gamma + \Lambda,
\end{align}

where the left-hand term represents advection and the three right-hand terms represent adiabatic cooling, radiative heating, and radiative cooling.  We plot all four terms in Figure \ref{fig:thermal_terms} for the Z=1 and Z=100 simulations.  At Z=1, adiabatic cooling is more important than radiative cooling at nearly all radii; at Z=100, radiative cooling is more important than adiabatic below 4 $R_p$.

We plot the radiative heating and cooling rates in Figure \ref{fig:heating_cooling_contributions}, which reveals a complex interplay between different processes.  In the solar metallicity model, cooling is dominated by free-free emission below 1.6 $R_p$, hydrogen line emission between 2 and 2.9 $R_p$, and recombination after 3.5 $R_p$.  Metal cooling still matters in this model--Fe II compromises 30\% of cooling from 1.6--2.0 $R_p$ and 2.9--3.4 $R_p$, for example--but hydrogen is clearly the most important coolant.  As the metallicity increases, metals become more and more prominent coolants.  The cooling in the 100$\times$ solar metallicity model is dominated by a diverse array of metal lines, including Na I, Ca II, Mg II, O III, and Fe VI, among many others.  In the solar metallicity model, the heating at radii beyond 1.3 $R_p$ is dominated by a combination of photoionization of HI, HeI, and (beyond 9 $R_p$) HeII; below 1.3 $R_p$, it is dominated by a combination of line heating and photoionization of metals like Si I.  The same pattern holds for the $10\times$ solar metallicity model.  At a metallicity of 30$\times$ solar, HI photoionization is only the most significant heating contribution from 1.9 to 4.5 $R_p$.  At a metallicity of 100$\times$ solar, it is only the most significant contributor from 1.7 to 2.4 $R_p$, with a diverse array of metals in various ionization states (like O II and Si I) dominating at larger radii.

These changes in heating and cooling cause changes in the temperature profile shown in Figure \ref{fig:excess_tpci}.  In the solar metallicity model, the temperature profile peaks at $2.49 R_p$ and 10,000 K, declining to 2800 K at $15 R_p$.  As the metallicity increases, the peak moves to lower temperatures and larger radii, while the temperature in the outer regions increases.  By the time the metallicity reaches 100$\times$ solar, the temperature peak has become more of an asymptote, as the temperature rises to 8000 K at 6 $R_p$ and then stays nearly constant beyond that point.  Consistent with this behavior, the total non-adiabatic cooling is significantly lower than the total heating in the solar metallicity model, but the two become comparable in the outer regions ($>6 R_p$) in the highest metallicity model.

\subsection{3D Microthena hydrodynamic models}\label{subsec:3d_models}
Our 3D models utilize the approach outlined in \citet{wang_2018}, \citet{wang_2020}, and \citet{zhang_2021}, which combines ray-tracing radiative transfer, real-time non-equilibrium thermochemistry, and hydrodynamics based on the higher-order Godunov method code \verb|Athena++| \citep{stone_2020}.  As in \cite{zhang_2021}, we perform radiative transfer by dividing photons into seven energy bins for the ray-tracing calculation: 1.4, 7, 12, 16, 47, 300, and 3000 eV, meant to represent IR/optical/NUV, FUV, Lyman-Werner photons (which photodissociate molecular hydrogen), soft EUV, hard EUV, soft X-rays, and hard X-rays respectively.  

We first run the model for the fiducial parameters: a planet mass of 9 $M_\Earth$ (consistent with the RV mass and mass-radius relation) and radius 2.9 $R_\Earth$, the stellar spectrum derived in Section \ref{subsec:stellar_spec}, a solar metallicity atmosphere, and a stellar wind with a solar velocity (400 km/s) and an 11$\times$ solar mass loss rate (see Section \ref{subsec:stellar_wind}).  We then selectively vary every parameter except the planet mass in an attempt to find a best fit model.  Microthena takes about one day to run a model on a GPU cluster, so running more than a handful of models is infeasible.

\begin{figure}
    \subfigure {\includegraphics[width=0.5\textwidth]{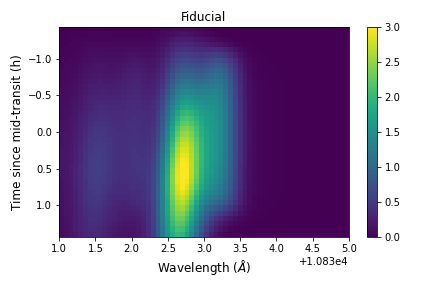}}
    \subfigure {\includegraphics[width=0.5\textwidth]{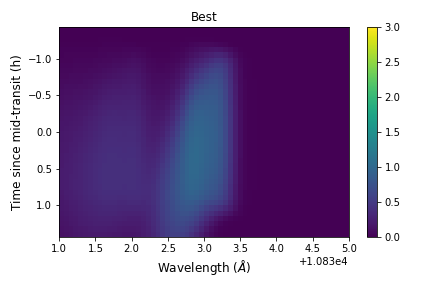}}
    \subfigure {\includegraphics[width=0.5\textwidth]{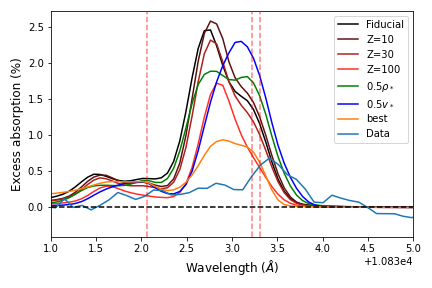}}
    \caption{Excess absorption as a function of wavelength and time, as predicted by the fiducial model (top) and the best-fit model (middle).  The fiducial model uses our best guess for all parameters, while the best-fit model has a higher stellar wind speed of 500 km/s (versus the fiducial 400 km/s), a higher metallicity (10$\times$ solar versus the fiducial 1$\times$ solar), and 1/3 the fiducial stellar EUV flux. Bottom: average in-transit excess absorption predicted by all Microthena models we tried, compared to data.}
    \label{fig:microthena_model_abs}
\end{figure}

\begin{figure*}
  \centering 
    \subfigure {\includegraphics
    [width=0.33\textwidth]{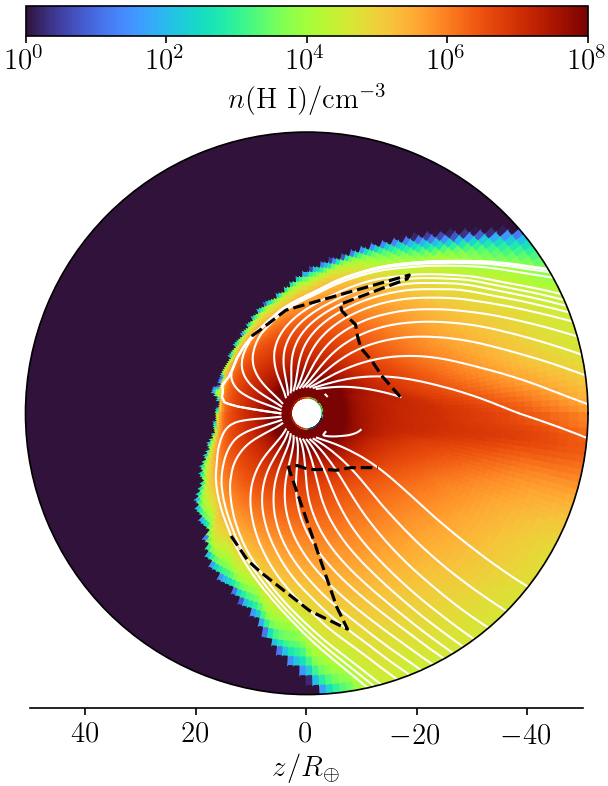}}\subfigure {\includegraphics
    [width=0.33\textwidth]{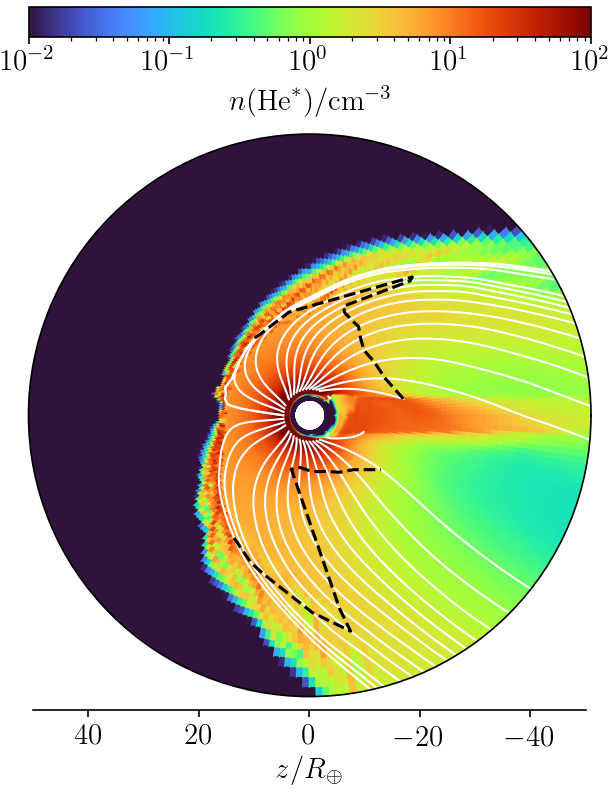}}\subfigure {\includegraphics
    [width=0.33\textwidth]{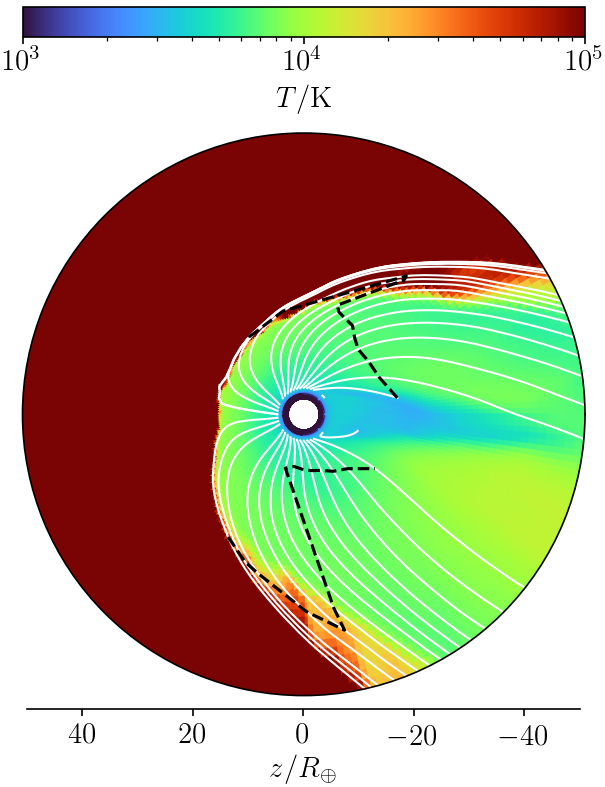}}
    \subfigure {\includegraphics
    [width=0.33\textwidth]{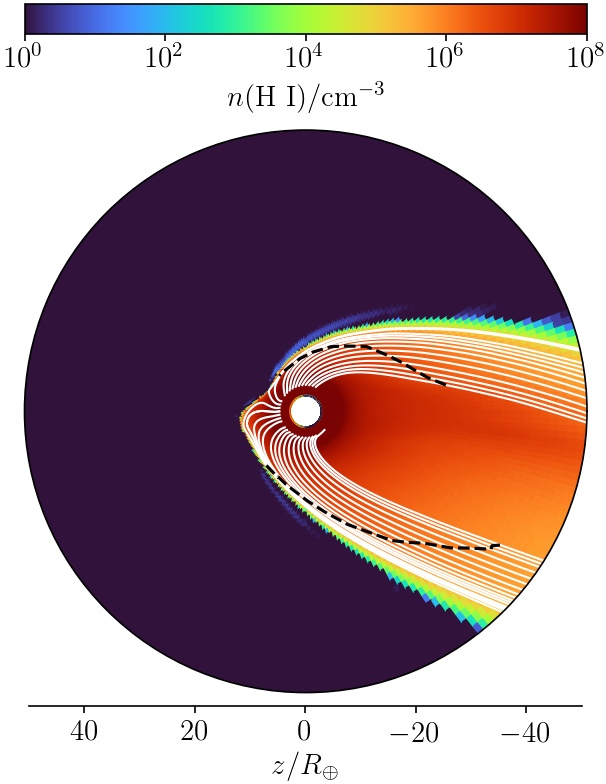}}\subfigure {\includegraphics
    [width=0.33\textwidth]{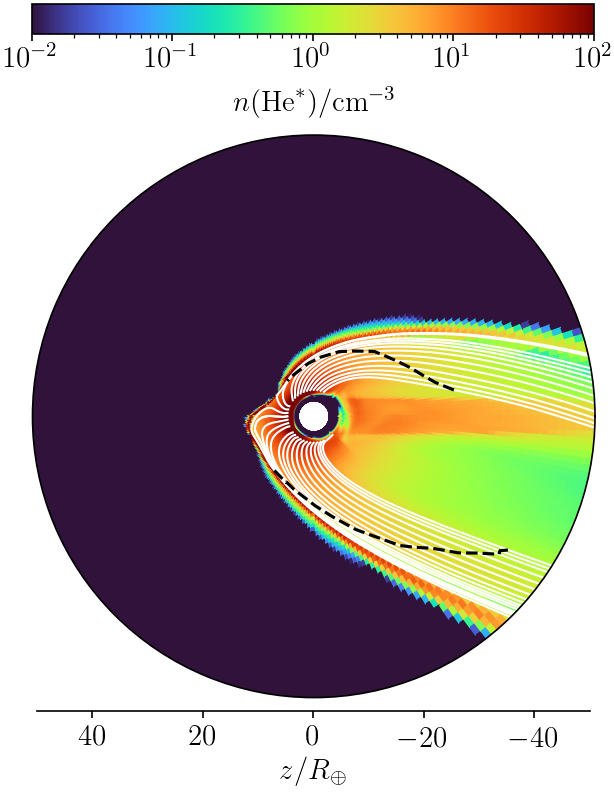}}\subfigure {\includegraphics
    [width=0.33\textwidth]{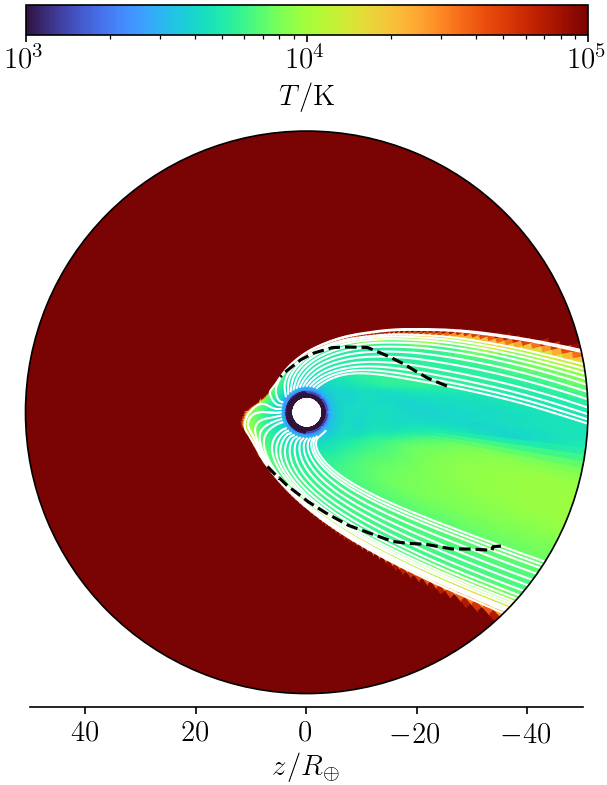}}
    \caption{Neutral hydrogen number density (left), triplet helium density (middle), and temperature (right) from the time-averaged (over the last $\sim 10$ kinematic timescales) fiducial (top) and best-fit (bottom) 3D models. The star is toward the left, and orbital motion is upwards. These plots show the profiles in the orbital plane. The white lines are the streamlines, while the dashed black lines represent the inner sonic surface.}
    \label{fig:slices}
\end{figure*}

The top panel of Figure \ref{fig:microthena_model_abs} shows the predicted absorption from the fiducial model with respect to time and wavelength, while the bottom panel compares the predicted average in-transit excess absorption spectrum to the observations.  This fiducial model significantly over-predicts the observed signal, with a predicted peak three times higher than the measured value.  The model absorption peak is also blueshifted by 15 km/s, while the measured peak is redshifted by $\sim$4 km/s.  Lastly, the fiducial model outflow contains two components, while the observed signal appears to be unimodal.  It is particularly difficult to explain the large discrepancy between the predicted and measured peak locations, as velocity blueshifts are a near-universal feature of all 3D outflow models \citep{wang_2018,wang_2020}.  If this 15 km/s discrepancy between model and data is due to a non-zero orbital eccentricity, it would be in severe tension with the radial velocity data, which indicates the planet is moving toward us at $7.4_{-7.0}^{+9.3}$ km/s at mid-transit--that is, in the wrong direction.  Interestingly, while the amplitude of the simulated signal is much too big, the band-integrated light curve has a similar shape to the observed light curve (see Figure \ref{fig:helium_lc}).  Both model and data exhibit a slow increase in absorption from ingress to 0.5 h past the white light mid-transit, followed by a fast decrease.

The poor match between the fiducial model and the data prompted us to try different combinations of metallicity, stellar wind speed, and stellar wind density. Figure \ref{fig:microthena_model_abs} shows the excess absorption spectrum of the different models we tried, compared to each other and to the data, while Table \ref{table:models_summary} reports their mass loss rates, peak absorption, and equivalent width of the absorption.  The qualitative behavior of the model when the metallicity is increased is very similar to the behavior of TPCI: the strength of helium absorption is similar at Z=1, 10, and 30, but substantially lower at Z=100.  In both models, absorption changes non-monotonically with metallicity, reaching its peak at Z=10.  The 3D models reveal that metallicity only marginally affects the blueshift.  In addition to increasing the metallicity, we tried halving the stellar wind density, which resulted in broader but slightly less blueshifted absorption with a slightly lower (by $\sim$30\%) peak.  We also tried halving the stellar wind speed, which marginally decreased the absorption peak and equivalent width, but decreased the blueshift by 10 km/s.  For all of these models, the mass loss rate remains remarkably similar: it decreases slightly with metallicity, from 0.12 $M_\Earth$/Gyr at Z=1 to 0.086 $M_\Earth$/Gyr at Z=100, but the fiducial, $0.5\rho_*$, and $0.5v_*$ models have mass loss rates within several percent of each other.  

Because all of the models that assume the fiducial stellar spectrum substantially overpredict the observed absorption, we also tried rescaling the input EUV flux.  Since the outflow is driven largely by EUV heating, and triplet helium is mostly produced by EUV-driven photoionization followed by recombination of electrons with helium, lowering the EUV flux should lower the corresponding absorption signal.  In the end, the model we found that best matches the observations (labelled ``Best'' in all the plots) has 1/3 the nominal EUV, a stellar wind with a speed of 500 km/s, and a metallicity of 10x.  As a result of the low EUV flux, the mass loss rate is 1/3 fiducial (Table \ref{table:models_summary}); as a result of the low mass loss rate and the higher stellar wind speed, the outflow is more confined.  In the middle and lower panels of Figure \ref{fig:microthena_model_abs}, we show results for this best fitting model.  This model is a much better match to the amplitude of the observed absorption signal and the corresponding transit light curve, although the predicted absorption signal is still bluer than the observed, and stronger in both the peak and the equivalent width.

Figure \ref{fig:slices} shows the neutral hydrogen number density, triplet helium number density, and temperature in the orbital plane for the fiducial and the best-fit models.    Both models exhibit a cometary tail, which is caused by the dense and fast stellar wind pushing the outflow away from the star.  The tail is angled partly because of the planetary orbital velocity, and partly because of the Coriolis force.  In both models, the stellar wind is strong enough to prevent the planetary outflow from becoming supersonic at many locations, making it possible for interplanetary conditions to affect the launching of the wind.  As expected, the ``best'' model has a substantially more confined outflow. 

\begin{table}[ht]
  \centering
  \caption{Summary of Model Predictions}
  \begin{tabular}{c c c c c}
  \hline
  	  Dim & Model & $\dot M$ ($M_\Earth$/Gyr) & Peak (\%) & EW (m\ang)\\
      \hline
      1 & Z=1 & 0.32 & 4.3 & 39.6 \\
      1 & Z=10 & 0.42 & 6.1 & 47.0 \\
      1 & Z=30 & 0.37 & 5.1 & 35.0\\
      1 & Z=100 & 0.13 & 1.7 & 9.6\\
      3 & \textbf{Fiducial} & 0.12 & 2.5 & 23.3 \\
      3 & Z=10 & 0.11 & 2.6 & 23.0 \\
      3 & Z=30 & 0.10 & 2.3 & 19.7\\
      3 & Z=100 & 0.086 & 1.7 & 12.5\\
      3 & 0.5$\rho_*$ & 0.12 & 1.9 & 22.3\\
      3 & 0.5$v_*$ & 0.12 & 2.3 & 20.5\\
      3 & \textbf{Best} & 0.041 & 0.9 & 11.4\\
      3 & Data & ? & $0.68 \pm 0.08$ & $7.0 \pm 0.4$\\
      \hline
  \end{tabular}
  \label{table:models_summary}
\end{table}

\subsection{Order-of-magnitude empirical estimate of mass loss rate}
By making a few assumptions, we can use the equivalent width of the helium signal to obtain a decent estimate for the mass of metastable helium in the outflow and an order-of-magnitude calculation of the mass loss rate.  The first assumption we make is that helium absorption from most of the planetary outflow is optically thin.  This is supported by the fact that we do not see the secondary peak at 10832 \ang, which is 1/8 the height of the primary peak in the optically thin regime, but higher when the primary peak becomes optically thick.  The optical thinness assumption tends to cause us to underestimate the mass loss rate.

The optical depth from the star to the observer is:

\begin{align}
    \tau(\lambda) = N\sigma_\lambda P(\lambda),
\end{align}
where N is the column density, $\sigma_\lambda \equiv (\pi e^2 g_lf_l)/(m_e c^2)$, and $P(\lambda)$ is the line profile ($\int_{-\infty}^{+\infty} P(\lambda) d\lambda = 1$).  Integrating $1 - e^{-\tau}$ over $\lambda$, and assuming the optically thin limit (where $1 - e^{-\tau} \approx \tau$), we obtain the standard equation for the equivalent width:

\begin{align}
    W_\lambda = N\sigma_\lambda
\end{align}

We can perform another integration over the cross-sectional area of the star, which turns the column density N into the total number of metastable helium atoms, N'.  We then divide each side by the cross-sectional area, so that the left hand side becomes the average equivalent width across the star--the number measured by our observations:

\begin{align}
    \iint W_\lambda dS &= \iint N\sigma_\lambda dS\\
    \frac{\iint W_\lambda dS}{\iint dS} &= N' \sigma_\lambda\\
    W_{avg} &= \frac{N' \sigma_\lambda}{\pi R_*^2}\\
    N' &= \frac{R_*^2 m_e c^2}{e^2 g_l f_l \lambda_0^2} W_{avg}\\
        &= 8 \times 10^{31}\\
    m_{He^*} &= 5 \times 10^8 g,
\end{align}
where we combined $g_l f_l$ for the two lines at 10833 \ang.  $5 \times 10^8$ g is 500 tons--a remarkably small amount of material to be detectable at interstellar distances.  This is the most secure part of our estimate.

Next, we convert the mass of metastable helium to total mass, by assuming that metastable helium comprises $10^{-6}$ of helium nuclei, and that helium is 25\% of total mass.  $10^{-6}$ is roughly the ratio we find in our 3D simulations from 2--5 $R_p$, as well as the ratio we find from 2--10 $R_p$ for HD 63433 b/c \citep{zhang_2021}.  We obtain $m_{\rm tot} = 2 \times 10^{15}$ g.

Finally, we estimate the replenishment lifetime of the metastable helium atoms: how long do they take to cross the stellar disk and stop being observable?  We calculate $\tau = R_* / c_s = 50,000$ s, adopting 10 km/s for the sound speed $c_s$.  The mass loss rate is then $m_{\rm tot} / \tau = 4 \times 10^{10} g/s = 0.22 M_\Earth/Gyr$.

This estimate should not be considered accurate to more than an order of magnitude, but it is reassuring that our empirical estimate is of the same order of magnitude as that of the simulations (Table \ref{table:models_summary}).

\section{Discussion}
\label{sec:discussion}
The fiducial versions of both our TPCI (1D) and Microthena (3D) models predict helium absorption many times stronger than observed.  For the same input parameters, the helium absorption predicted by TPCI is much higher than that predicted by Microthena (4.3\% vs. 2.5\%).  The 3D structure, which TPCI cannot model, determines the shape of the helium absorption signal in both wavelength and time.  Though the predicted helium signal is too large, the predicted mass loss rates in Table \ref{table:models_summary} are all reasonable.  We assume a 2\% hydrogen/helium envelope, which is consistent with the planet's radius, mass, and equilibrium temperature assuming a rocky core \citep{zeng_2008}.  The mass loss timescales predicted by Microthena would then range from 1.6--2.4 Gyr, and those predicted by TPCI would range from 0.4--1.6 Gyr.  All of these timescales are comparable to the planet age, which is high enough that the planet has likely lost a substantial portion of its primordial envelope, but low enough that it is not surprising the planet still has an envelope.

We next consider whether there might be additional modifications to the 3D Microthena models that could better match the amplitude of the observed signal while also reducing the size of the predicted blueshift.  We could increase the stellar wind speed by a larger amount, which would eventually succeed in suppressing the mass loss rate \citep{murray-clay_2009}, but this would also increase the magnitude of the predicted blueshift.  If we instead decreased the EUV flux by a larger factor than 3x, it would also decrease the helium signal strength, but our X-ray observations strongly disfavor this scenario.  The scaling relations from \cite{linsky_2014} that we used to estimate the EUV spectrum have a dispersion about the fit line of 20--37\% for F5--K5 stars, which seems to make a 3x error implausible, but the relations are based on the Ly$\alpha$ flux, which we have not measured for TOI 560 and must infer from the X-ray flux and the stellar rotation rate.  In addition, the dispersion might be smaller for young stars only than for all stars.  The X-ray and MUV (1230--2588 \ang) fluxes are even better known because they were directly measured by XMM-Newton within weeks of both Keck observations.  A substantially larger planetary mass (e.g. 16 $M_\Earth$) could also suppress the signal, but would be in tension with both the RV measurements and the mass-radius relation of \cite{chen_2016}.  This mass-radius relation is derived primarily from observations of mature planets; since young planets are expected to be more inflated, it provides a conservative upper limit on the likely mass of this planet.  Finally, the redshift would be more consistent with the models if we assumed a much weaker wind, in which case there would be a significant up-orbit stream launched from the dayside, as seen in the 3D hydrodynamic simulations by \cite{mccann_2019}.  This is scenario favored by \cite{czesla_2021} for the redshift they observe from HAT-P-32b; however, their helium signal also shows pre-ingress absorption, which ours does not.

Our 1D TPCI models demonstrate that increasing the metallicity beyond 10$\times$ solar gives rise to a rich tapestry of radiative phenomena, which in turn results in significant changes to the structure of the outflow and a reduction in the predicted helium absorption strength.  Our 3D models show a qualitatively similar pattern.  It is clear, however, that further work is required in order to better address some of the limitations of our models for describing metal-rich atmospheres.  For example, both codes are potentially inaccurate at high metallicities because they use the escape probability formalism for radiative transfer, instead of performing radiative transfer completely correctly.  The 3D Microthena model only accounts for line cooling from O, S, Si, O+, S+, and Si+, whereas our TPCI models predict that there is a much broader array of elements and ionization states that are important for both heating and cooling at high metallicities (Figure \ref{fig:heating_cooling_contributions}).  In addition, while the 3D model divides all radiation into 7 bins, TPCI uses the incident spectrum at the resolution provided (Figure \ref{fig:stellar_spectrum}); this likely has a significant impact on the predicted thermal structure of the outflow.  

In addition to ensuring correct behavior of the model at high metallicities, future studies of metal-rich outflows should also vary the assumed elemental ratios (e.g., C/O) to reflect the diverse array of possible atmospheric compositions for sub-Neptune-sized exoplanets.  Condensation of species like NaCl, MgSiO$_4$, and TiO in the deep atmosphere could also further alter the composition of the outflow in ways that may be detectable in future observations.

The parameter space exploration we undertook assumes that the observed outflow is in a steady state.  However, there is some tentative evidence for variability in our data: on the second night, the absorption during the portion of transit we could observe was slightly stronger and slightly bluer than on the first night.  This effect is only marginally significant (2-3$\sigma$), and could be explained by a combination of statistical fluctuations, underestimated errors, stellar variability, and unknown instrumental systematics.  However, our 3D simulations show that stellar wind conditions have a significant impact on the absorption signal, so it is by no means implausible that variations in the stellar wind could lead to variations in the outflow properties.  During the 2008 solar minimum, for example, the solar wind number density regularly fluctuated by a factor of two over a timescale of days, with dozens of upward spikes corresponding to increases of more than a factor of five \citep{lei_2011}.  The solar wind velocity during this period regularly fluctuated between 400 and 600 km/s, again on a timescale of days.  TOI 560 is younger, less massive, and more active than the Sun, and we do not know where it was in its activity cycle during our observations.  It is therefore possible that TOI 560's stellar wind may be more variable than that of the Sun, and that the stellar wind conditions during the two transits were appreciably different.  Our 3D simulations show that decreasing the stellar wind density by a factor of two decreases the absorption peak by $\sim$30\%, while decreasing the wind speed by a factor of two decreases the blueshift by $\sim10$ km/s.  This level of variation is similar to the magnitude of the observed variability between the two nights.

Another effect that neither TPCI nor Microthena accounts for is magnetic fields.  Using the same methodology as \cite{zhang_2020}, which is itself based on \cite{owen_2014}, we can estimate the significance of magnetic fields by comparing the magnetic pressure $P_B = B^2/(8 \pi)$ to the ram pressure $P_{\rm ram} = \dot{M} v / (8 \pi r^2)$.  At 3.6 $R_p$, roughly the position of the helium line photosphere, our Microthena models predict a typical wind speed of 7 km/s.  As discussed in \cite{zhang_2020}, the slow rotation of a tidally locked planet (the tidal circulation timescale is 1200 yr, assuming Q=100) gives rise to a weak magnetic field, which does not significantly affect the outflow ($P_{\rm ram} / P_B = 64$ at 3.6 $R_p$).  However, the interplanetary magnetic field can plausibly have a significant effect on the outflow.  Based on the scaling relation between surface magnetic field and stellar age in \cite{vidotto_2014}, we calculate an interplanetary magnetic field of 0.025 G at the position of TOI 560.01, which leads to $P_{\rm ram} / P_B = 1.1$ at 3.6 $R_p$.  Finally, we emphasize that these are order-of-magnitude calculations with uncertainties of at least a factor of a few in all quantities: interplanetary magnetic fields, stellar activity cycles, and planetary magnetic fields are all very poorly understood for extrasolar systems.

\section{Conclusion}
\label{sec:conclusion}
We observed two transits of TOI 560.01 using Keck/NIRSPEC and detected a strong helium absorption signal during each transit.  The observed signal shows a number of intriguing features, including a slow rise in absorption accompanied by a fast decline, a slight redshift, and tentative evidence of variability between the two transits, with the signal being stronger and bluer on the second night at 2-3$\sigma$ significance.  TOI 560.01 is the first mini Neptune with a helium detection.  Its size and youth place it in the critical regime where mini Neptunes transition into super Earths, giving rise to the radius gap and dramatically shaping exoplanet demographics.

We applied the 1D code TPCI and the 3D code Microthena to model the outflow.  We found it difficult to match the low magnitude of the observed absorption level and the moderate redshift with a solar metallicity planetary atmosphere.  Increasing the metallicity to Z=100 suppresses the helium absorption signal in both TPCI and Microthena, but in both models, intermediate metallicities do not.  However, neither code is very well suited to modeling high metallicity atmospheres.  This problem is more acute for Microthena, which uses a simplified radiative transfer model and includes a limited number of atomic coolants.  Further work will be necessary to confirm whether a high metallicity atmosphere is a good explanation for the data, and if so, whether it is the only plausible explanation.

As a low mass mini Neptune around a nearby K star, TOI 560.01 is a favorable target for atmospheric characterization.  In this paper, we measure the planet's helium absorption, but its Ly$\alpha$ absorption may also be observable with HST/STIS.  If so, it will provide complementary insight into the escaping atmosphere: due to the strength of the Ly$\alpha$ line and the strong interstellar absorption, which wipes out the core, Ly$\alpha$ traces energetic hydrogen atoms in the tenuous outer regions of the outflow.  JWST could also reveal important insights into the outflow by measuring the composition and thermal structure of the deeper atmosphere, which, aside from being scientifically valuable in its own right, will drastically shrink the parameter space of inputs that models can assume.

Finally, TOI 560 is a two planet system, and TOI 560.02 is also a transiting mini Neptune.  This makes the system an excellent test for mass loss models.  The two planets share the same contemporary X-ray/EUV environment, and the same irradiation history.  In addition, planets of similar size located in adjacent orbits might be expected to have largely similar formation and/or migration histories, and therefore it is reasonable to expect that their primordial atmospheric compositions would be quite similar.  This is supported by observational studies of the masses and radii of multi-planet systems in the Kepler sample, which suggest that planets in the same system tend to have similar masses and radii (the `peas in a pod' theory; \citealt{weiss_2018}).

\textit{Software:}  \texttt{numpy \citep{van_der_walt_2011}, scipy \citep{virtanen_2020}, matplotlib \citep{hunter_2007}, dynesty \citep{speagle_2019}, SAS}

\acknowledgments
The helium data presented herein were obtained at the W. M. Keck Observatory, which is operated as a scientific partnership among the California Institute of Technology, the University of California and the National Aeronautics and Space Administration. The Observatory was made possible by the generous financial support of the W. M. Keck Foundation.

Based on observations obtained with XMM-Newton, an ESA science mission with instruments and contributions directly funded by ESA Member States and NASA. 

\bibliographystyle{apj} \bibliography{main}

\begin{thebibliography}{64}
\expandafter\ifx\csname natexlab\endcsname\relax\def\natexlab#1{#1}\fi

\bibitem[{{Allart} {et~al.}(2019){Allart}, {Bourrier}, {Lovis}, {Ehrenreich},
  {Aceituno}, {Guijarro}, {Pepe}, {Sing}, {Spake}, \&
  {Wyttenbach}}]{allart_2019}
{Allart}, R., {Bourrier}, V., {Lovis}, C., {et~al.} 2019, \aap, 623, A58

\bibitem[{{Barrag{\'a}n} {et~al.}(2021){Barrag{\'a}n}, {Armstrong}, {Gandolfi},
  {Carleo}, {Vidotto}, {Villarreal D'Angelo}, {Oklop{\v{c}}i{\'c}}, {Isaacson},
  {Oddo}, {Collins}, {Fridlund}, {Sousa}, {Persson}, {Hellier}, {Howell},
  {Howard}, {Redfield}, {Eisner}, {Georgieva}, {Dragomir}, {Bayliss},
  {Nielsen}, {Klein}, {Aigrain}, {Zhang}, {Teske}, {Twicken}, {Jenkins},
  {Esposito}, {Van Eylen}, {Rodler}, {Adibekyan}, {Alarcon}, {Anderson}, {Akana
  Murphy}, {Barrado}, {Barros}, {Benneke}, {Bouchy}, {Bryant}, {Butler},
  {Burt}, {Cabrera}, {Casewell}, {Chaturvedi}, {Cloutier}, {Cochran}, {Crane},
  {Crossfield}, {Crouzet}, {Collins}, {Dai}, {Deeg}, {Deline}, {Demangeon},
  {Dumusque}, {Figueira}, {Furlan}, {Gnilka}, {Goad}, {Goffo},
  {Guti{\'e}rrez-Canales}, {Hadjigeorghiou}, {Hartman}, {Hatzes}, {Harris},
  {Henderson}, {Hirano}, {Hojjatpanah}, {Hoyer}, {Kab{\'a}th}, {Korth},
  {Lillo-Box}, {Luque}, {Marmier}, {Mo{\v{c}}nik}, {Muresan}, {Murgas},
  {Nagel}, {Osborne}, {Osborn}, {Osborn}, {Palle}, {Raimbault}, {Ricker},
  {Rubenzahl}, {Stockdale}, {Santos}, {Scott}, {Schwarz}, {Shectman},
  {Raimbault}, {Seager}, {S{\'e}gransan}, {Serrano}, {Skarka}, {Smith},
  {{\v{S}}ubjak}, {Tan}, {Udry}, {Watson}, {Wheatley}, {West}, {Winn}, {Wang},
  {Wolfgang}, \& {Ziegler}}]{barragan_2021}
{Barrag{\'a}n}, O., {Armstrong}, D.~J., {Gandolfi}, D., {et~al.} 2021, arXiv
  e-prints, arXiv:2110.13069

\bibitem[{{Bourrier} {et~al.}(2013){Bourrier}, {Lecavelier des Etangs},
  {Dupuy}, {Ehrenreich}, {Vidal-Madjar}, {H{\'e}brard}, {Ballester},
  {D{\'e}sert}, {Ferlet}, {Sing}, \& {Wheatley}}]{bourrier_2013}
{Bourrier}, V., {Lecavelier des Etangs}, A., {Dupuy}, H., {et~al.} 2013, \aap,
  551, A63

\bibitem[{{Bourrier, V.} {et~al.}(2018){Bourrier, V.}, {Lecavelier des Etangs,
  A.}, {Ehrenreich, D.}, {Sanz-Forcada, J.}, {Allart, R.}, {Ballester, G. E.},
  {Buchhave, L. A.}, {Cohen, O.}, {Deming, D.}, {Evans, T. M.}, {Garc\'{\i}a
  Mu\~noz, A.}, {Henry, G. W.}, {Kataria, T.}, {Lavvas, P.}, {Lewis, N.},
  {L\'opez-Morales, M.}, {Marley, M.}, {Sing, D. K.}, \& {Wakeford, H.
  R.}}]{bourrier_2018b}
{Bourrier, V.}, {Lecavelier des Etangs, A.}, {Ehrenreich, D.}, {et~al.} 2018,
  A\&A, 620, A147

\bibitem[{Chen \& Kipping(2016)}]{chen_2016}
Chen, J., \& Kipping, D. 2016, The Astrophysical Journal, 834, 17

\bibitem[{{Czesla} {et~al.}(2021){Czesla}, {Lamp{\'o}n}, {Sanz-Forcada},
  {Garc{\'\i}a Mu{\~n}oz}, {L{\'o}pez-Puertas}, {Nortmann}, {Yan}, {Nagel},
  {Yan}, {Schmitt}, {Aceituno}, {Amado}, {Caballero}, {Casasayas-Barris},
  {Henning}, {Khalafinejad}, {Molaverdikhani}, {Montes}, {Pall{\'e}},
  {Reiners}, {Schneider}, {Ribas}, {Quirrenbach}, {Zapatero Osorio}, \&
  {Zechmeister}}]{czesla_2021}
{Czesla}, S., {Lamp{\'o}n}, M., {Sanz-Forcada}, J., {et~al.} 2021, arXiv
  e-prints, arXiv:2110.13582

\bibitem[{{de Jager} {et~al.}(1966){de Jager}, {Namba}, \&
  {Neven}}]{de_jager_1966}
{de Jager}, C., {Namba}, O., \& {Neven}, L. 1966, \bain, 18, 128

\bibitem[{{Ferland} {et~al.}(2013){Ferland}, {Porter}, {van Hoof}, {Williams},
  {Abel}, {Lykins}, {Shaw}, {Henney}, \& {Stancil}}]{ferland_2013}
{Ferland}, G.~J., {Porter}, R.~L., {van Hoof}, P.~A.~M., {et~al.} 2013, \rmxaa,
  49, 137

\bibitem[{{Fulton} \& {Petigura}(2018)}]{fulton_2018}
{Fulton}, B.~J., \& {Petigura}, E.~A. 2018, \aj, 156, 264

\bibitem[{{Fulton} {et~al.}(2017){Fulton}, {Petigura}, {Howard}, {Isaacson},
  {Marcy}, {Cargile}, {Hebb}, {Weiss}, {Johnson}, {Morton}, {Sinukoff},
  {Crossfield}, \& {Hirsch}}]{fulton_2017}
{Fulton}, B.~J., {Petigura}, E.~A., {Howard}, A.~W., {et~al.} 2017, \aj, 154,
  109

\bibitem[{{Gaidos} {et~al.}(2020){Gaidos}, {Hirano}, {Wilson}, {France},
  {Rockcliffe}, {Newton}, {Feiden}, {Krishnamurthy}, {Harakawa}, {Hodapp},
  {Ishizuka}, {Jacobson}, {Konishi}, {Kotani}, {Kudo}, {Kurokawa}, {Kuzuhara},
  {Nishikawa}, {Omiya}, {Serizawa}, {Tamura}, {Ueda}, \&
  {Vievard}}]{gaidos_2020b}
{Gaidos}, E., {Hirano}, T., {Wilson}, D.~J., {et~al.} 2020, \mnras, 498, L119

\bibitem[{Gaidos {et~al.}(2020)Gaidos, Hirano, Mann, Owens, Berger, France,
  Vanderburg, Harakawa, Hodapp, Ishizuka, Jacobson, Konishi, Kotani, Kudo,
  Kurokawa, Kuzuhara, Nishikawa, Omiya, Serizawa, Tamura, \&
  Ueda}]{gaidos_2020a}
Gaidos, E., Hirano, T., Mann, A.~W., {et~al.} 2020, Monthly Notices of the
  Royal Astronomical Society, 495, 650

\bibitem[{{Ginzburg} {et~al.}(2018){Ginzburg}, {Schlichting}, \&
  {Sari}}]{ginzburg_2018}
{Ginzburg}, S., {Schlichting}, H.~E., \& {Sari}, R. 2018, \mnras, 476, 759

\bibitem[{{Gupta} \& {Schlichting}(2019)}]{gupta_2019}
{Gupta}, A., \& {Schlichting}, H.~E. 2019, \mnras, 487, 24

\bibitem[{{Hunter}(2007)}]{hunter_2007}
{Hunter}, J.~D. 2007, Computing in Science and Engineering, 9, 90

\bibitem[{{Husser} {et~al.}(2013){Husser}, {Wende-von Berg}, {Dreizler},
  {Homeier}, {Reiners}, {Barman}, \& {Hauschildt}}]{husser_2013}
{Husser}, T.-O., {Wende-von Berg}, S., {Dreizler}, S., {et~al.} 2013, \aap,
  553, A6

\bibitem[{{Kasper} {et~al.}(2020){Kasper}, {Bean}, {Oklop{\v{c}}i{\'c}},
  {Malsky}, {Kempton}, {D{\'e}sert}, {Rogers}, \& {Mansfield}}]{kasper_2020}
{Kasper}, D., {Bean}, J.~L., {Oklop{\v{c}}i{\'c}}, A., {et~al.} 2020, \aj, 160,
  258

\bibitem[{{Lavie} {et~al.}(2017){Lavie}, {Ehrenreich}, {Bourrier}, {Lecavelier
  des Etangs}, {Vidal-Madjar}, {Delfosse}, {Gracia Berna}, {Heng}, {Thomas},
  {Udry}, \& {Wheatley}}]{lavie_2017}
{Lavie}, B., {Ehrenreich}, D., {Bourrier}, V., {et~al.} 2017, \aap, 605, L7

\bibitem[{{Lee} \& {Connors}(2021)}]{lee_2021}
{Lee}, E.~J., \& {Connors}, N.~J. 2021, \apj, 908, 32

\bibitem[{{Lei} {et~al.}(2011){Lei}, {Thayer}, {Wang}, \&
  {McPherron}}]{lei_2011}
{Lei}, J., {Thayer}, J.~P., {Wang}, W., \& {McPherron}, R.~L. 2011, \solphys,
  274, 427

\bibitem[{{Linsky} {et~al.}(2014){Linsky}, {Fontenla}, \&
  {France}}]{linsky_2014}
{Linsky}, J.~L., {Fontenla}, J., \& {France}, K. 2014, \apj, 780, 61

\bibitem[{{Linsky} {et~al.}(2013){Linsky}, {France}, \& {Ayres}}]{linsky_2013}
{Linsky}, J.~L., {France}, K., \& {Ayres}, T. 2013, \apj, 766, 69

\bibitem[{{Mamajek} \& {Hillenbrand}(2008)}]{mamajek_2008}
{Mamajek}, E.~E., \& {Hillenbrand}, L.~A. 2008, \apj, 687, 1264

\bibitem[{{Mazeh} {et~al.}(2007){Mazeh}, {Tamuz}, \& {Zucker}}]{mazeh_2007}
{Mazeh}, T., {Tamuz}, O., \& {Zucker}, S. 2007, Astronomical Society of the
  Pacific Conference Series, Vol. 366, {The Sys-Rem Detrending Algorithm:
  Implementation and Testing}, ed. C.~{Afonso}, D.~{Weldrake}, \& T.~{Henning},
  119

\bibitem[{{McCann} {et~al.}(2019){McCann}, {Murray-Clay}, {Kratter}, \&
  {Krumholz}}]{mccann_2019}
{McCann}, J., {Murray-Clay}, R.~A., {Kratter}, K., \& {Krumholz}, M.~R. 2019,
  \apj, 873, 89

\bibitem[{Mignone {et~al.}(2007)Mignone, Bodo, Massaglia, Matsakos, Tesileanu,
  Zanni, \& Ferrari}]{mignone_2007}
Mignone, A., Bodo, G., Massaglia, S., {et~al.} 2007, The Astrophysical Journal
  Supplement Series, 170, 228

\bibitem[{{Mills} {et~al.}(2019){Mills}, {Howard}, {Petigura}, {Fulton},
  {Isaacson}, \& {Weiss}}]{mills_2019}
{Mills}, S.~M., {Howard}, A.~W., {Petigura}, E.~A., {et~al.} 2019, \aj, 157,
  198

\bibitem[{{Mills} \& {Mazeh}(2017)}]{mills_2017}
{Mills}, S.~M., \& {Mazeh}, T. 2017, \apjl, 839, L8

\bibitem[{{Mousis} {et~al.}(2020){Mousis}, {Deleuil}, {Aguichine}, {Marcq},
  {Naar}, {Aguirre}, {Brugger}, \& {Gon{\c{c}}alves}}]{mousis_2020}
{Mousis}, O., {Deleuil}, M., {Aguichine}, A., {et~al.} 2020, \apjl, 896, L22

\bibitem[{{Murray-Clay} {et~al.}(2009){Murray-Clay}, {Chiang}, \&
  {Murray}}]{murray-clay_2009}
{Murray-Clay}, R.~A., {Chiang}, E.~I., \& {Murray}, N. 2009, \apj, 693, 23

\bibitem[{{Niermann} {et~al.}(2010){Niermann}, {B{\"o}ke}, {Sadeghi}, \&
  {Winter}}]{niermann_2010}
{Niermann}, B., {B{\"o}ke}, M., {Sadeghi}, N., \& {Winter}, J. 2010, European
  Physical Journal D, 60, 489

\bibitem[{{Oklop{\v{c}}i{\'c}}(2019)}]{oklopcic_2019}
{Oklop{\v{c}}i{\'c}}, A. 2019, \apj, 881, 133

\bibitem[{{Oklop{\v{c}}i{\'c}} \& {Hirata}(2018)}]{oklopcic_2018}
{Oklop{\v{c}}i{\'c}}, A., \& {Hirata}, C.~M. 2018, \apj, 855, L11

\bibitem[{{Owen}(2019)}]{owen_2018}
{Owen}, J.~E. 2019, Annual Review of Earth and Planetary Sciences, 47, 67

\bibitem[{Owen \& Adams(2014)}]{owen_2014}
Owen, J.~E., \& Adams, F.~C. 2014, Monthly Notices of the Royal Astronomical
  Society, 444, 3761

\bibitem[{{Owen} \& {Wu}(2017)}]{owen_2017}
{Owen}, J.~E., \& {Wu}, Y. 2017, \apj, 847, 29

\bibitem[{{Palle} {et~al.}(2020){Palle}, {Nortmann}, {Casasayas-Barris},
  {Lamp{\'o}n}, {L{\'o}pez-Puertas}, {Caballero}, {Sanz-Forcada}, {Lara},
  {Nagel}, {Yan}, {Alonso-Floriano}, {Amado}, {Chen}, {Cifuentes},
  {Cort{\'e}s-Contreras}, {Czesla}, {Molaverdikhani}, {Montes}, {Passegger},
  {Quirrenbach}, {Reiners}, {Ribas}, {S{\'a}nchez-L{\'o}pez}, {Schweitzer},
  {Stangret}, {Zapatero Osorio}, \& {Zechmeister}}]{palle_2020}
{Palle}, E., {Nortmann}, L., {Casasayas-Barris}, N., {et~al.} 2020, \aap, 638,
  A61

\bibitem[{{Pequignot} {et~al.}(1991){Pequignot}, {Petitjean}, \&
  {Boisson}}]{pequignot_1991}
{Pequignot}, D., {Petitjean}, P., \& {Boisson}, C. 1991, \aap, 251, 680

\bibitem[{{Salz} {et~al.}(2015{\natexlab{a}}){Salz}, {Banerjee}, {Mignone},
  {Schneider}, {Czesla}, \& {Schmitt}}]{salz_2015}
{Salz}, M., {Banerjee}, R., {Mignone}, A., {et~al.} 2015{\natexlab{a}}, \aap,
  576, A21

\bibitem[{{Salz} {et~al.}(2016){Salz}, {Czesla}, {Schneider}, \&
  {Schmitt}}]{salz_2016}
{Salz}, M., {Czesla}, S., {Schneider}, P.~C., \& {Schmitt}, J.~H.~M.~M. 2016,
  \aap, 586, A75

\bibitem[{{Salz} {et~al.}(2015{\natexlab{b}}){Salz}, {Schneider}, {Czesla}, \&
  {Schmitt}}]{salz_2015b}
{Salz}, M., {Schneider}, P.~C., {Czesla}, S., \& {Schmitt}, J.~H.~M.~M.
  2015{\natexlab{b}}, \aap, 576, A42

\bibitem[{{Salz} {et~al.}(2018){Salz}, {Czesla}, {Schneider}, {Nagel},
  {Schmitt}, {Nortmann}, {Alonso-Floriano}, {L{\'o}pez-Puertas}, {Lamp{\'o}n},
  {Bauer}, {Snellen}, {Pall{\'e}}, {Caballero}, {Yan}, {Chen}, {Sanz-Forcada},
  {Amado}, {Quirrenbach}, {Ribas}, {Reiners}, {B{\'e}jar}, {Casasayas-Barris},
  {Cort{\'e}s-Contreras}, {Dreizler}, {Guenther}, {Henning}, {Jeffers},
  {Kaminski}, {K{\"u}rster}, {Lafarga}, {Lara}, {Molaverdikhani}, {Montes},
  {Morales}, {S{\'a}nchez-L{\'o}pez}, {Seifert}, {Zapatero Osorio}, \&
  {Zechmeister}}]{salz_2018}
{Salz}, M., {Czesla}, S., {Schneider}, P.~C., {et~al.} 2018, \aap, 620, A97

\bibitem[{{Schlaufman}(2010)}]{schlaufman_2010}
{Schlaufman}, K.~C. 2010, \apj, 719, 602

\bibitem[{{Smette} {et~al.}(2015){Smette}, {Sana}, {Noll}, {Horst}, {Kausch},
  {Kimeswenger}, {Barden}, {Szyszka}, {Jones}, {Gallenne}, {Vinther},
  {Ballester}, \& {Taylor}}]{smette_2015}
{Smette}, A., {Sana}, H., {Noll}, S., {et~al.} 2015, \aap, 576, A77

\bibitem[{{Spake} {et~al.}(2018){Spake}, {Sing}, {Evans}, {Oklop{\v{c}}i{\'c}},
  {Bourrier}, {Kreidberg}, {Rackham}, {Irwin}, {Ehrenreich}, {Wyttenbach},
  {Wakeford}, {Zhou}, {Chubb}, {Nikolov}, {Goyal}, {Henry}, {Williamson},
  {Blumenthal}, {Anderson}, {Hellier}, {Charbonneau}, {Udry}, \&
  {Madhusudhan}}]{spake_2018}
{Spake}, J.~J., {Sing}, D.~K., {Evans}, T.~M., {et~al.} 2018, \nat, 557, 68

\bibitem[{Speagle(2020)}]{speagle_2019}
Speagle, J.~S. 2020, Monthly Notices of the Royal Astronomical Society, 493,
  3132

\bibitem[{{Stone} \& {Proga}(2009)}]{stone_2009}
{Stone}, J.~M., \& {Proga}, D. 2009, \apj, 694, 205

\bibitem[{{Stone} {et~al.}(2020){Stone}, {Tomida}, {White}, \&
  {Felker}}]{stone_2020}
{Stone}, J.~M., {Tomida}, K., {White}, C.~J., \& {Felker}, K.~G. 2020, \apjs,
  249, 4

\bibitem[{{Turon} {et~al.}(1993){Turon}, {Creze}, {Egret}, {Gomez}, {Grenon},
  {Jahrei{\ss}}, {Requieme}, {Argue}, {Bec-Borsenberger}, {Dommanget},
  {Mennessier}, {Arenou}, {Chareton}, {Crifo}, {Mermilliod}, {Morin},
  {Nicolet}, {Nys}, {Prevot}, {Rousseau}, {Perryman}, {Arlot}, {Baglin},
  {Barthes}, {Baylac}, {Brosche}, {Burnet}, {Delhaye}, {Dettbarn}, {Erbach},
  {Figueras}, {Fricke}, {Helmer}, {Hemenway}, {Jordi}, {Lampens}, {Lederle},
  {Lub}, {Manfroid}, {Mattci}, {Mazurier}, {Mermilliod}, {Morrison}, {Murray},
  {Oblak}, {Perie}, {Pernier}, {Le Poole}, {Quijano}, {Rapaport}, {Sellier},
  {Torra}, {Tucholke}, {de Vegt}, {Argyle}, {Bacchus}, {Baron}, {Calaf},
  {Cordoni}, {Fabricius}, {Feaugas}, {Fehlberg}, {Florkowski}, {de Geus},
  {Gibbs}, {Hartmann}, {Jauncey}, {Johnston}, {Marouard}, {Mekkas}, {Muinos},
  {Nunez}, {Ochsenbein}, {de Orus}, {Paredes}, {Penston}, {Petersen}, {Peyrin},
  {Robin}, {Roman}, {Rossello}, {Schwan}, {Sinachopoulos}, {White},
  {Zacharias}, {Hog}, {Kovalevsky}, {van Leeuwen}, {Lindegren}, {Schutz}, \&
  {Schrijver}}]{turon_1993}
{Turon}, C., {Creze}, M., {Egret}, D., {et~al.} 1993, Bulletin d'Information du
  Centre de Donnees Stellaires, 43, 5

\bibitem[{{van der Walt} {et~al.}(2011){van der Walt}, {Colbert}, \&
  {Varoquaux}}]{van_der_walt_2011}
{van der Walt}, S., {Colbert}, S.~C., \& {Varoquaux}, G. 2011, Computing in
  Science and Engineering, 13, 22

\bibitem[{{Van Eck} {et~al.}(2017){Van Eck}, {Neyskens}, {Jorissen}, {Plez},
  {Edvardsson}, {Eriksson}, {Gustafsson}, {J{\o}rgensen}, \&
  {Nordlund}}]{van_eck_2017}
{Van Eck}, S., {Neyskens}, P., {Jorissen}, A., {et~al.} 2017, \aap, 601, A10

\bibitem[{{Vidal-Madjar} {et~al.}(2008){Vidal-Madjar}, {Lecavelier des Etangs},
  {D{\'e}sert}, {Ballester}, {Ferlet}, {H{\'e}brard}, \&
  {Mayor}}]{vidal-madjar_2008}
{Vidal-Madjar}, A., {Lecavelier des Etangs}, A., {D{\'e}sert}, J.~M., {et~al.}
  2008, \apjl, 676, L57

\bibitem[{{Vidotto} {et~al.}(2014){Vidotto}, {Gregory}, {Jardine}, {Donati},
  {Petit}, {Morin}, {Folsom}, {Bouvier}, {Cameron}, {Hussain}, {Marsden},
  {Waite}, {Fares}, {Jeffers}, \& {do Nascimento}}]{vidotto_2014}
{Vidotto}, A.~A., {Gregory}, S.~G., {Jardine}, M., {et~al.} 2014, \mnras, 441,
  2361

\bibitem[{{Virtanen} {et~al.}(2020){Virtanen}, {Gommers}, {Oliphant},
  {Haberland}, {Reddy}, {Cournapeau}, {Burovski}, {Peterson}, {Weckesser},
  {Bright}, {van der Walt}, {Brett}, {Wilson}, {Jarrod Millman}, {Mayorov},
  {Nelson}, {Jones}, {Kern}, {Larson}, {Carey}, {Polat}, {Feng}, {Moore}, {Vand
  erPlas}, {Laxalde}, {Perktold}, {Cimrman}, {Henriksen}, {Quintero}, {Harris},
  {Archibald}, {Ribeiro}, {Pedregosa}, {van Mulbregt}, \&
  {Contributors}}]{virtanen_2020}
{Virtanen}, P., {Gommers}, R., {Oliphant}, T.~E., {et~al.} 2020, Nature
  Methods, 17, 261

\bibitem[{Wang \& Dai(2018)}]{wang_2018}
Wang, L., \& Dai, F. 2018, The Astrophysical Journal, 860, 175

\bibitem[{{Wang} \& {Dai}(2020)}]{wang_2020}
{Wang}, L., \& {Dai}, F. 2020, arXiv e-prints, arXiv:2101.00045

\bibitem[{{Wang} \& {Dai}(2021)}]{wang_2021}
---. 2021, \apj, 914, 99

\bibitem[{{Weiss} {et~al.}(2018){Weiss}, {Marcy}, {Petigura}, {Fulton},
  {Howard}, {Winn}, {Isaacson}, {Morton}, {Hirsch}, {Sinukoff}, {Cumming},
  {Hebb}, \& {Cargile}}]{weiss_2018}
{Weiss}, L.~M., {Marcy}, G.~W., {Petigura}, E.~A., {et~al.} 2018, \aj, 155, 48

\bibitem[{{Wood} \& {Linsky}(2010)}]{wood_2010}
{Wood}, B.~E., \& {Linsky}, J.~L. 2010, \apj, 717, 1279

\bibitem[{{Wood} {et~al.}(2005){Wood}, {M{\"u}ller}, {Zank}, {Linsky}, \&
  {Redfield}}]{wood_2005b}
{Wood}, B.~E., {M{\"u}ller}, H.~R., {Zank}, G.~P., {Linsky}, J.~L., \&
  {Redfield}, S. 2005, \apjl, 628, L143

\bibitem[{{Wood} {et~al.}(2021){Wood}, {M{\"u}ller}, {Redfield}, {Konow},
  {Vannier}, {Linsky}, {Youngblood}, {Vidotto}, {Jardine},
  {Alvarado-G{\'o}mez}, \& {Drake}}]{wood_2021}
{Wood}, B.~E., {M{\"u}ller}, H.-R., {Redfield}, S., {et~al.} 2021, \apj, 915,
  37

\bibitem[{{Zeng} \& {Seager}(2008)}]{zeng_2008}
{Zeng}, L., \& {Seager}, S. 2008, \pasp, 120, 983

\bibitem[{{Zhang} {et~al.}(2021{\natexlab{a}}){Zhang}, {Knutson}, {Wang},
  {Dai}, {Oklopcic}, \& {Hu}}]{zhang_2020}
{Zhang}, M., {Knutson}, H.~A., {Wang}, L., {et~al.} 2021{\natexlab{a}}, \aj,
  161, 181

\bibitem[{{Zhang} {et~al.}(2021{\natexlab{b}}){Zhang}, {Knutson}, {Wang},
  {Dai}, {dos Santos}, {Fossati}, {Henry}, {Ehrenreich}, {Alibert}, {Hoyer},
  {Wilson}, \& {Bonfanti}}]{zhang_2021}
---. 2021{\natexlab{b}}, arXiv e-prints, arXiv:2106.05273

\end{thebibliography}

\end{document}